\documentclass[a4paper,twocolumn]{aa}
\usepackage[]{natbib}
\usepackage[]{multirow}
\usepackage{graphicx}
\usepackage[a4paper]{hyperref}
\idline{}{1}
\usepackage{booktabs}
\usepackage{longtable}
\usepackage{txfonts}
\usepackage{lscape}
\usepackage{rotating}
\usepackage{epsf}
\usepackage[]{multirow}
\usepackage{graphicx}

\newcommand{\be}{\begin{equation}}
\newcommand{\ee}{\end{equation}}
\newcommand{\ba}{\begin{array}{ll}}
\newcommand{\ea}{\end{array}}

\newcommand{\BSAX}{{\em Beppo}SAX}

\newcommand{\BWFC}{{\em Beppo}SAX WFC}
\newcommand{\bc}{\begin{center}}
\newcommand{\ec}{\end{center}}

\def\ergcms{\rm erg\,cm$^{-2}$\,s$^{-1}$}
\def\ergcmsk{erg\,cm$^{-2}$\,s$^{-1}$\,keV$^{-1}$}
\def\ergcm{\rm erg\,cm^{-2}}

\def\ctss{\rm cts\,s^{-1}}
\def\k{\rm \,keV}

\def\dtnumproc{7295} %

\def\fstnumproc{228}  %
\def\colcolsam{7529}
\def\colsam{7439}

\def\dtnum{8283} %
\def\fstnum{253} %

\def \AAP #1 #2 {{\em Astron. Astrophys.\/} {\bf #1}, #2}
\def \AAL #1 #2 {{\em Astron. Astrophys. Lett.\/} {\bf #1}, L#2}
\def \AAR #1 #2 {{\em Astron. Astrophys. Rev.\/} {\bf #1}, #2}
\def \AAS #1 #2 {{\em Astron. Astrophys. Suppl. Ser.\/} {\bf #1}, #2}
\def \AJ #1 #2 {{\em Astron. J.\/} {\bf #1}, #2}
\def \ANNREV #1 #2 {{\em Ann. Rev. Astron. Astrophys.\/} {\bf #1}, #2}
\def \APJ #1 #2 {{\em Astrophys. J.\/} {\bf #1}, #2}
\def \APJL #1 #2 {{\em Astrophys. J. Lett.\/} {\bf #1}, L#2}
\def \APJS #1 #2 {{\em Astrophys. J. Suppl.\/} {\bf #1}, #2}
\def \APSS #1 #2 {{\em Astrophys. Space Sci.\/} {\bf #1}, #2}
\def \ASR #1 #2 {{\em Adv. Space Res.\/} {\bf #1}, #2}
\def \BAIC #1 #2 {{\em Bull. Astron. Inst. Czechosl.\/} {\bf #1}, #2}
\def \JSQRT #1 #2 {{\em J. Quant. Spectrosc. Radiat. Transfer\/} {\bf #1}, #2}
\def \MN #1 #2 {{\em Mon. Not. R. Astr. Soc.\/} {\bf #1}, #2}
\def \MEM #1 #2 {{\em Mem. R. Astr. Soc.\/} {\bf #1}, #2}
\def \PLR #1 #2 {{\em Phys. Lett. Rev.\/} {\bf #1}, #2}
\def \PASJ #1 #2 {{\em Publ. Astron. Soc. Japan\/} {\bf #1}, #2}
\def \PASP #1 #2 {{\em Publ. Astr. Soc. Pacific\/} {\bf #1}, #2}
\def \NAT #1 #2 {{\em Nature\/} {\bf #1}, #2}
\def \SAIT #1 #2 {{\em Mem.\ Soc.\ Astron.\ It.\/} {\bf #1}, #2}
\def \MESS #1 #2 {{\em The Messenger\/} {\bf #1}, #2}
\def \ASTRNACH #1 #2 {{\em Astron. Nach.\/} {\bf #1}, #2}
\input epsf.sty
\begin{document}
\bibliographystyle{astron}
\author{F.\,Verrecchia\inst{1},
    J.\,J.\,M.\,in 't Zand\inst{2},
    P.\,Giommi\inst{1},
    P.\,Santolamazza\inst{1},
    S.\,Granata\inst{1},
    J.\,J.\,Schuurmans\inst{2} and
    L.\,A.\,Antonelli\inst{1,3}
}

\title{The \BWFC\ X--ray source catalogue}

\institute{
ASI Science Data Center ({\em ASDC})$^{*}$, c/o ESA-ESRIN, via Galileo Galilei, I--00044 Frascati, Rome, Italy
\and
 Netherlands Institute for Space Reasearch ({\em SRON}), Sorbonnelaan 2, 3584, CA Utrecht, The Netherlands
\and
INAF -- Osservatorio Astronomico di Roma ({\em OAR}), via Frascati 33, Monteporzio Catone, Rome, Italy
}
\date{Received: 29 December 2006 / Accepted: 26 June 2007}

\offprints{verrecchia@asdc.asi.it \\ $^{*}$ INAF-OAR personnel resident at ASDC under ASI contract I/024/05}
\authorrunning{Verrecchia et al.}
\titlerunning{The \BWFC\ source catalogue}

\abstract{}
{We present the catalogue of X--ray sources detected by the two Wide Field Cameras (WFCs) in complete observations
on board \BSAX\ during its 6 years of operational lifetime, between April 1996 and April 2002.}
{The \BSAX\ WFCs were coded mask instruments sensitive in the 2--28\,\k\ energy band with a 40x40 square degree fields of
view, pointing in opposite directions and perpendicularly to the \BSAX\ Narrow Field Instruments (NFI).
 The WFCs were usually operated simultaneously to NFI observations, each lasting up to several days. WFCs observed thus the 
entire sky several times with a typical sensitivity of 2 to 10 mCrab.
 A systematic analysis of all WFC observations in the \BSAX\ archive has been carried out 
using the latest post-mission release of the WFC analysis software and calibrations.
}
{The catalogue includes \fstnum\ distinct sources, obtained from a total sample of \dtnum\ WFC detections. We describe the basic statistical properties of the sample and present a six-year history of two celestial calibration X--ray sources.}
{} 
\maketitle

\section{Introduction}

The \BSAX\ X--ray astronomy satellite (\citealt*{Boe}), was a major program of the Italian
Space Agency (ASI) with participation of the Netherlands Agency for Aereospace Programs (NIVR).
The main scientific goal of the mission was to perform spectroscopic
and timing studies of several classes of X--ray sources in a very broad energy band
(0.1--300\,\k).
The satellite was launched on April 30, 1996 and observations were carried out until April 30, 2002.

\begin{figure*}[ht]

    \centering
    \resizebox{\hsize}{!}{\rotatebox[]{-90}{\includegraphics[width=1.2\textwidth]{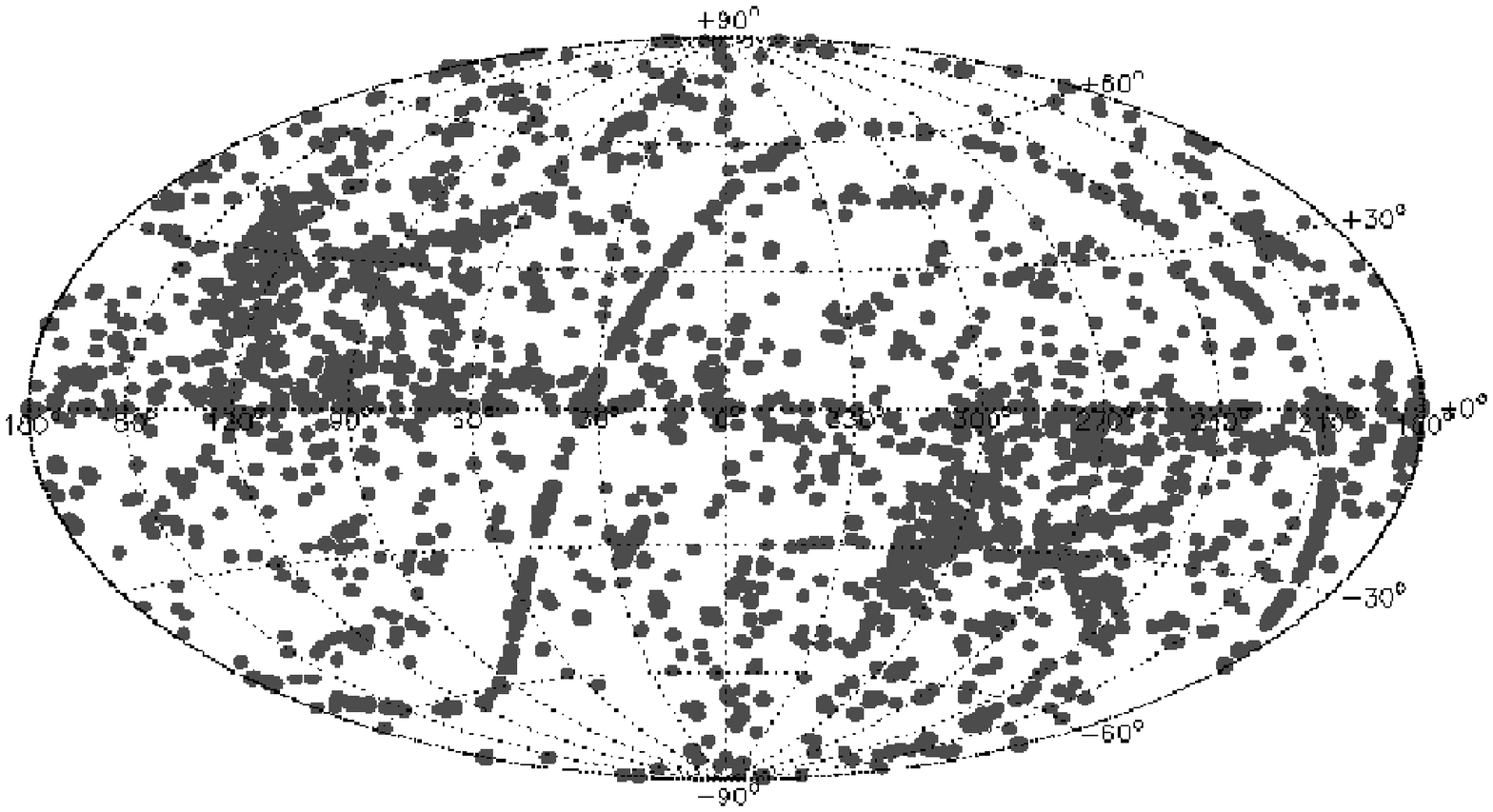}}  

    \rotatebox[]{-90}{\includegraphics[width=1.1\textwidth]{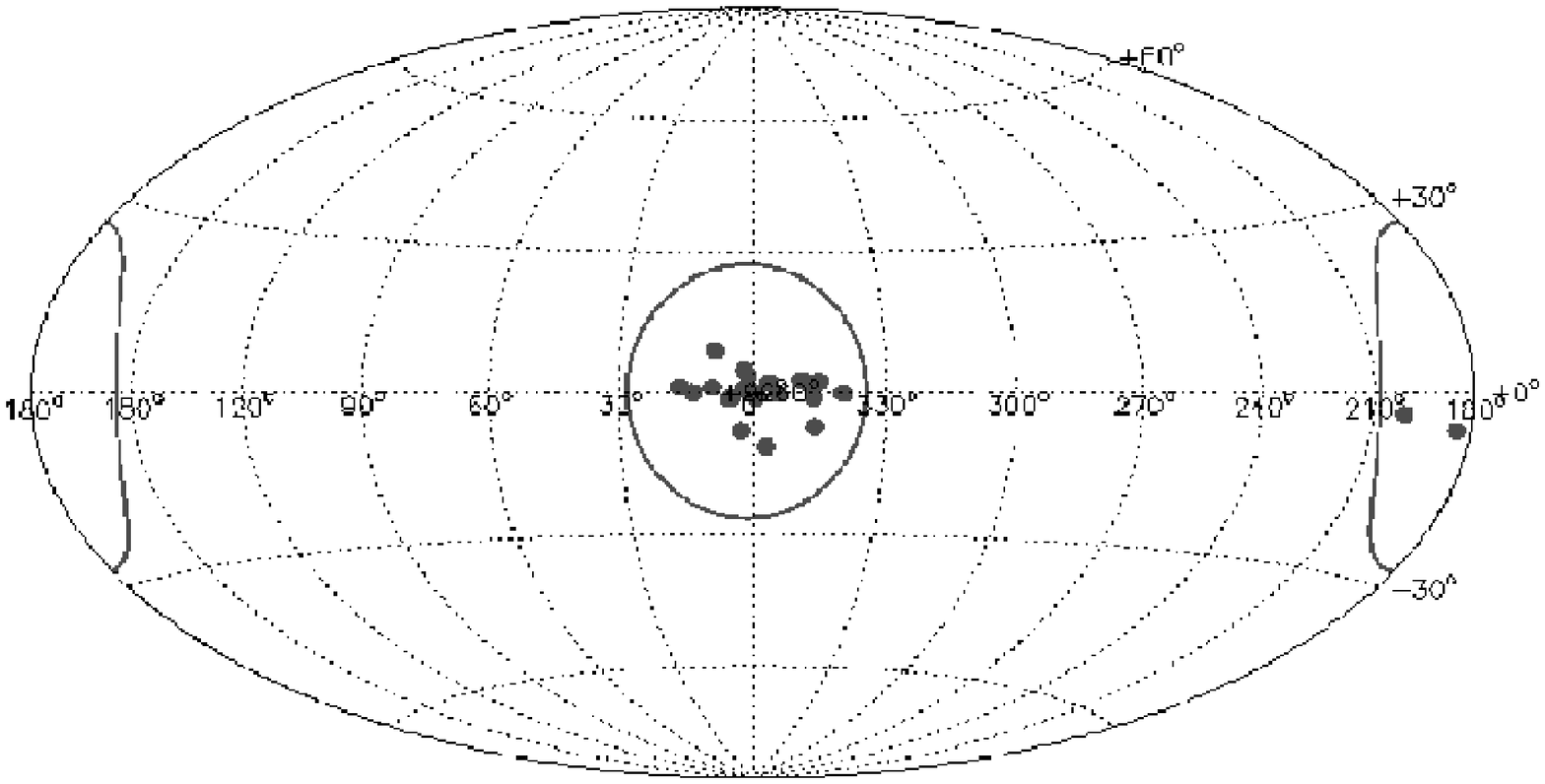}}
}
\vspace{-2cm}

   \caption{The Aitoff projection in galactic coordinates of all the WFC observation pointing directions (left) and of the FOV of a typical WFC pointing of the Galactic Center (the circle includes the squared WFC FOV). WFC unit 1 points at Sgr-A and unit 2 at the anti-center, including the Crab Nebula.
		}
	 \label{Fig0}
   \end{figure*}

The scientific payload comprised four Narrow Field Instruments (NFIs) and two Wide Field Cameras (WFCs) 
pointing in opposite directions from each other and perpendicular to the NFI (\citealt*{Boe}; \citealt*{Jag}).
The WFC observations covered the whole sky multiple times over the period 1996-2002 (see Fig. \ref{Fig0}).	
\\

The WFCs large field of view (FOV) and good sensitivity made these instruments suitable to detect and locate X--ray transient sources with a precision of a few arcminutes and to monitor large parts of the sky. Among the most important results obtained with the WFCs is the discovery of the X--ray afterglow phenomenon of Gamma Ray Bursts in 1997 (\citealt*{Costa}). Moreover, the 12 monitoring campaigns of the Galactic Bulge allowed the spectral and timing monitoring of known persistent and transient X--ray sources and the discovery of many new galactic X--ray transients and
phenomena (see for example \citealt*{Ube99}; \citealt*{Corn00}; \citealt*{Zand01} and 2004a; \citealt*{Corn03}). These observations resulted in a significant advance in the knowledge of the transient emission properties of low mass X--ray binaries
(e.g., \citealt*{Zand04b}, 2004c).

 After the end of the mission a systematic data analysis of the entire WFCs archive was carried out.
This analysis, based on the processing of two complementary datasets, allowed us to build the  
complete catalogue of WFC X--ray sources and an archive of high-level data products, including energy spectra and light curves 
of the brighter sources. This archive is accessible on the web through the ASDC multi-mission interactive archive
(http://www.asdc.asi.it/mmia/).
%
%

In this paper we present the final \BWFC\ source catalogue including \fstnum\
distinct and previously known X--ray sources except for a new transient. The catalogue was obtained from a total sample of \dtnum\ true detections.

 This catalogue is the first complete all-sky survey above 2.5\,\k\ since {\em HEAO-1 A4} (\citealt*{HEAO})
with 6 years of coverage corresponding to a net total exposure of over 100 million seconds and with a far better location accuracy. It contains more information on variability over long time scale in the hard X--ray band than for example the ROSAT all-sky survey catalogues, the ASCA 0.7--10\,\k\ (\citealt*{ASCA}) and the first three INTEGRAL-IBIS low galactic latitude surveys (\citealt*{INTS}, \citealt*{INTS2}, \citealt*{INTH}) and the high galactic latitude survey of the Swift-BAT (\citealt*{BAT}), and comparable with the last INTEGRAL-IBIS survey (\citealt*{INTS3}).

\section{Instrument and software description}

\subsection{General characteristics and the raw data archive}

The WFCs were two identical coded mask instruments for the imaging of the hard X--ray sky (1.8--28\,\k; \citealt*{Jag}) developed and built by SRON.
Their FOV was $40^\circ$\,$\times$\,$40^\circ$ each (full width to zero response) with an angular resolution of 5\arcmin\ (full width at half maximum, FWHM); the spectral and time resolution were 20\% at 6\,\k\ (FWHM) and 0.5\,{\rm ms}, respectively. 
 The source location accuracy for bright sources (for the standard attitude reconstruction from star sensor data) was 1.4\arcmin\ at 99\% confidence level, and larger for less significant sources as measured through the signal-to-noise ratio (SNR). The on-axis sensitivity was in the range 2--10 {\rm mCrab} for a typical \BWFC\ observation of 3\,$\times$\,10$^4$\,{\rm s}, and depended on the intensities of other sources in the same FOV.

These properties made the WFCs suitable instruments for the monitoring of medium to high intensity X--ray sources and for the study of transient X-ray phenomena.\\

 The WFCs raw data archive has been stored in 2652 Final Observation Tapes (FOTs). FOTs corresponding to contiguous observing periods with the same pointing and roll angle, were combined to form a multi-FOT complete observation.

 We carried out two data processing runs on two different data samples. The first pertains to all data when no part of the FOV is occulted by the Earth. Some FOTs were, thus, completely discarded and 1438 observations remained. The second 
sample pertains to all data (1543 observations), whether occulted by the Earth or not, and exposure corrections were applied for each source
 to take into account occultations. We performed two different processing runs because the first allowed a straightforward modelling of the detector background image resulting in less image artifacts, while the second allowed to obtain the longest source exposure to reduce statistical noise and increase source time coverage.

\begin{table*}[!ht]
\centering
\caption{\footnotesize{The astronomical databases used to build the reference catalogue. Database acronyms, objects class and the references for each catalogue are shown from left to right. GC and GP are the Galactic Center and Plane.}}
\begin{tabular}{llccc}
\hline \hline

 Database Name~~~& Objects class & References  \\

\hline
 XRB         & X-ray Binary systems & \citealt*{XRB}  \\
 HMXRB & High Mass X-ray Binary systems & \citealt*{HMXB}    \\
 LMXRB & Low Mass X-ray Binary systems & \citealt*{LMXB}   \\
 INTEGRAL-IBIS GP Soft $\gamma$-ray Catalogue & X-ray Binary systems & \citealt*{INTS}   \\
 INTEGRAL-IBIS Hard X-ray GC Survey  &  & \citealt*{INTH} \\
 Swift-BAT High Galactic Latitude survey & AGN & \citealt*{BAT}  \\
 ZCAT        & Galaxies with B magnitude $<$\,14 & \citealt*{Zcat}  \\
 AGN         & Active Galactic Nuclei & \citealt*{Pad97} and  \\
   &     & private communication    \\
 RASS-BSC   & ROSAT All-Sky Survey Bright Sources & \citealt*{Vog99}   \\
    & with 0.1--2.4\,\k\ flux $>$\,10$^{-11}$\,{\rm erg s}$^{-1}${\rm cm}$^{-2}$ &     \\
 Sedentary   &  High Energy Peaked BL LACs (HBL) & \citealt*{Gio99}, \citealt*{Gio05}   \\
 CABSCAT     & Spectroscopic binary systems & \citealt*{Str88}, \citealt*{Str93}  \\
 CVCAT       & Cataclismic variable stars & \citealt*{CVcat}   \\
 ALLWARPS    & Cluster of galaxies & \citealt*{AllWarps}  \\
 BASSANI     & Absorbed active galactic nuclei & private communication   \\

\hline
\end{tabular}
\label{Cats}

\end{table*}

\subsection{The WFC imaging algorithm}

The WFCs are shadow mask cameras (\citealt*{Ables}; \citealt*{Dick68}).
 An opaque screen with a pseudo-random pattern of holes (coded mask) is placed in front of a position-sensitive detector (a multi-wire proportional counter; \citealt*{Mels88}), so that X--ray sources in the FOV cast shadows on the detector, each displaced according to the off-axis position of the source. The WFC masks and detectors were of the same size and, therefore, only the on-axis position is fully coded with the mask pattern. However this configuration allows a better angular resolution at the same FOV compared to systems having the entire FOV fully coded.

 The sky image is reconstructed with an iterative cleaning algorithm ('Iterative Removal Of Sources', IROS; \citealt*{Hamm86}, \citealt*{Hamm}; \citealt*{Zand92}), based on a cross-correlation of the detector image with the coded mask pattern (\citealt*{Fen78}).
 IROS performs a new cross-correlation during each iteration and new detections are localised via a fit with the expected Point Spread Function (PSF) and then compared to the positions of X--ray sources in a reference catalogue. Finally, each source that is significantly above the noise, is simulated on the detector plane, including its energy and angle-dependent spatial response, and then subtracted.

The main systematic error in the reconstructed sky is due to a residual spatial non-linearity in the detectors which cannot be accounted for.
As a result physical pixel sizes are not completely identical and consequently the PSF shows inaccuracies. The main impact is on the sensitivity in very crowded FOVs for exposures longer than a few days (e.g., the Galactic Center field, hereafter GC) and on the source 
location accuracy which is, in this case, hard limited to a best-case value of 1.4\arcmin\ (99\% confidence level; \citealt*{Hei}).

\subsection{The reference catalogue}

The reference catalogue of X--ray sources used in the first data processing run includes 17674 entries. It was built combining various tables of known astronomical objects (see Table \ref{Cats}) in order to include a large sample of entries from extra-galactic catalogues to allow the detection of AGN and cluster of galaxies in the 2--10\,\k\ band at a flux level near the limiting sensitivity. Each WFC FOV includes from a few hundred up to two thousand sources of the reference catalogue. This requires an adjustment of the detection threshold criteria since source positions are more likely to coincide with a noise peak when there are many ``potential'' sources in the FOV.
 In the second data processing run a smaller reference catalogue was used to reduce the probability
of a coincidence of a known source with a noise peak. It was built by selecting sources with 1\k\ flux density greater than 4.8\,$\times$ 10$^{-12}$\,\ergcmsk\ and including validated sources from the first processing run.

\subsection{WFC processing procedure}

Raw data consist of a list of events labelled with detector position,
energy channel and detection time. Fluxes and positions of sources detected in an observation are determined from the raw data in three main steps (\citealt*{Jag}):
\begin{itemize}
\item {\em I}: a 'cleaned' event list is obtained from each FOT for time intervals with known and stable attitude, excluding all unwanted periods (such as those due to abnormal values of housekeeping parameters and passages through the South Atlantic Geomagnetic Anomaly). In the first processing run times were selected to keep the Earth outside the FOV, while in the second one periods were included when the Earth is inside the FOV (determining for each source the exposure when it is not occulted). Event lists from each FOT in a complete observation are merged;
\item {\em II}: data products such as detector and sky images, and fluxes in specific energy bands are generated from event lists through the decoding process. A first decoding run is executed in the most sensitive band (i.e. 2--8\,\k, assuming a Crab-like spectrum), to select significant sources. Then the decoding algorithm is executed in all energy bands for these sources. In the second processing run a further decoding test in the 8--19\,\k\ band was added to search for hard transient sources;
\item {\em III}: higher level scientific products, such as spectral fits with simple models, spectral energy distributions and light curves are generated in the same bands for each detected source, applying SNR acceptance thresholds of 4.5 and 6.5 for sparse and crowded fields respectively.\end{itemize}
The SNR is calculated by the WFC software using as noise the standard deviation evaluated on source plus background counts. This is because for the most part of the sources the standard deviation is background dominated. Thresholds were preliminarly defined taking into account the large reference catalogue and the sensitivity dependence on the combined intensity of all sources in the FOV (see detailed description in section 3.3).

 At the end of the standard procedure, a visual inspection of the scientific products (spectra and light curves) was performed.

\section{Results}

 In this section we describe how the fluxes were validated with data of well-documented stable X--ray sources and how the WFC catalogue was built through the creation of the final sample of detections.

\subsection{The observations of the Crab Nebula and Cas A}

 As validation of both the WFC software and the standard analysis procedure at high, medium and low flux levels, we checked all the observations of a sample of well-documented stable X--ray sources including the Crab Nebula, Cas A and the Tycho supernova remnant. We show here some results for the Crab and Cas A from the first data processing run.

\begin{figure}[!ht]
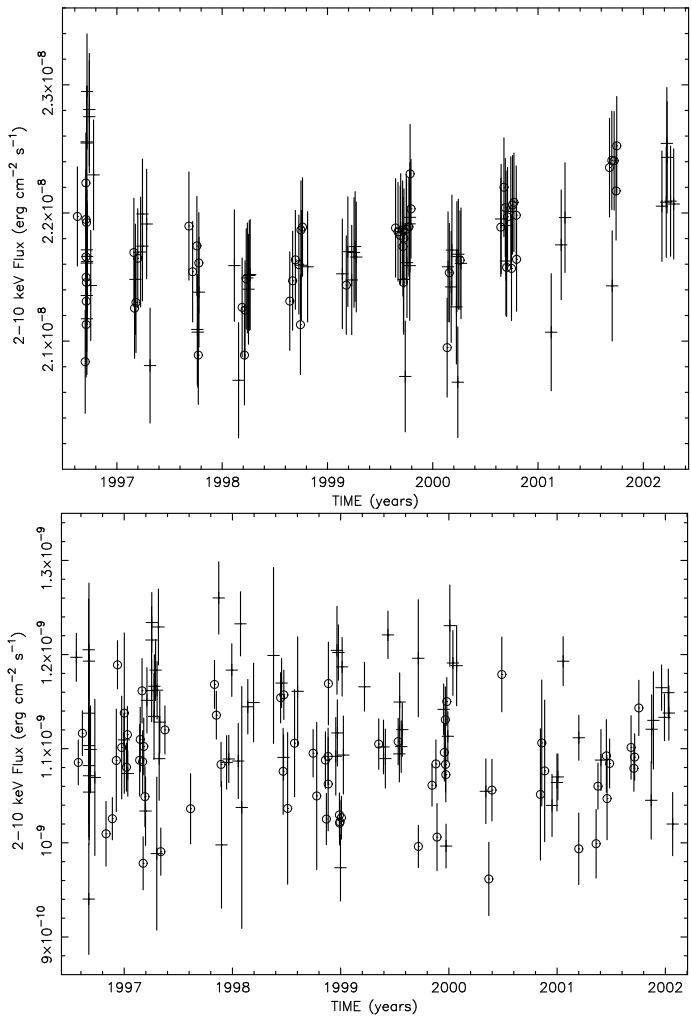

\centering
   \resizebox{\hsize}{!}{\rotatebox[]{-90}{
   \includegraphics[width=\textwidth ]{verrecchia_fig3.ps}
   \includegraphics[width=\textwidth ]{verrecchia_fig4.ps}
}}

     \caption[t]{
The 2-10\,\k\ Crab Nebula (top panel) and Cas A secular light curves corrected for the linear decrease with off-axis angle described in the text (open circles correspond to unit 1). A systematic error of $\sim$\,2\%\ was added in quadrature to the corrected flux errors for both WFCs to reach a reduced chi-square of $\sim$\,1. Residual dispersions for corrected light curves are 2\% and 6.2\% respectively.}
              \label{Fig3}%
\end{figure}

We analyzed 130 observations of the Crab Nebula at off-axis angles lower than \,21$\degr$ and exposures ranging between 1.67 and 129.7\,{\rm ks}. 
 We calculated the relative deviations from the Crab Nebula 2--10\,\k\ mean flux (2.12\,$\times$\,$10^{-8}$\,\ergcms, evaluated using the final WFC response matrices) and corrected fluxes for a residual linear trend with off-axis angle between FOV center and borders of about 7\%. This trend is mainly due to the decrease with the cosine of the angle of the number of photons detected per square cm. We plotted then the resulting 2--10\,keV secular light curve of Crab Nebula (Fig. \ref{Fig3}). A systematic error of about 2\,\% was added in quadrature to the flux errors to make the data consistent with an expected reduced chi-square of 1.
This error was applied to all the detections in the catalogue. We show in the bottom panel of Fig. \ref{Fig3} the Cas A corrected secular light curve obtained from 144 observations, compatible with a residual dispersion of $\sim$\,6.2\%. This dispersion is larger than the Crab one at the lower flux level of Cas A.
 A further systematic error of about 4.1\% is needed to obtain again a reduced chi-square of 1. This term has not been added to flux errors.
\begin{table}[!ht]

\centering \caption{\footnotesize{Number of sources and detections for the main source classes in the WFC catalogue. The table does not include 37 detections of 2 X--ray pulsar sources, 4U 0142+614 and PSR J0540-6919.}}
\begin{tabular}{lcc}
\hline \hline
 Source Class~~~ & Number of  & Number of   \\
  & detections & sources  \\
\hline
 candidate X--ray binaries & 2 & 2 \\
 High Mass X--ray binaries & 2045 & 49  \\
 Low Mass X--ray binaries & 4876 & 87  \\
 Cataclismic Variables & 125 & 11  \\
 Extended Galactic sources & 443 & 5  \\
 Stars & 99 & 21  \\
 White Dwarfs & 15 & 2  \\
 Galaxies & 14 & 2 \\
 AGN & 472 & 45  \\
 Cluster of Galaxies & 59 & 13  \\
 Unclassified & 96 & 14   \\
\hline

\hline
\end{tabular}
\label{Classes}
\end{table}

\subsection{Source identifications}

Sources in the reference catalogue (see section 2.3) were automatically identified by the IROS algorithm and then double-checked visually.
Sources not identified or misidentified in the reference catalogue were cross checked with literature regarding sources discovered by ASCA, RXTE, \BSAX, INTEGRAL, Swift-BAT, XMM (for example see \citealt*{ASCA}, \citealt*{Gio00}, \citealt*{INTS2} and 2007, \citealt*{BAT}, \citealt*{1XMM} and 2006). Moreover we checked and updated, when necessary, source classifications.

\subsection{Source detection filtering}

The sensitivity at a certain position in the FOV depends on the off-axis
angle (due to a dependence on effective area), on the particle-induced
and cosmic diffuse backgrounds and on the combined
intensity of all X-ray point sources in the FOV. The first three
components are rather stable under nominal conditions, the latter
varies considerably with galactic coordinates (the variations of the cosmic background above 2\k\ are much smaller; see for example \citealt*{ACXB} and \citealt*{XCXB}). The largest concentration 
of bright X-ray sources is close to the GC, where consequently the sensitivity is 
worse than elsewhere in the sky.

 In fields crowded with bright sources there is a complicating factor related to a non-optimum imaging by the
detectors. The detectors suffer from a differential non-linearity (\citealt*{Jag}). Due to the dependence
on the spectrum and off-axis angles of all detected sources, it is not
completely accounted for in the IROS algorithm. The effect is so strong that a
systematic sensitivity limit is reached for GC fields for an exposure
longer than a few hundred ks.

%
   \begin{figure*}[!ht]
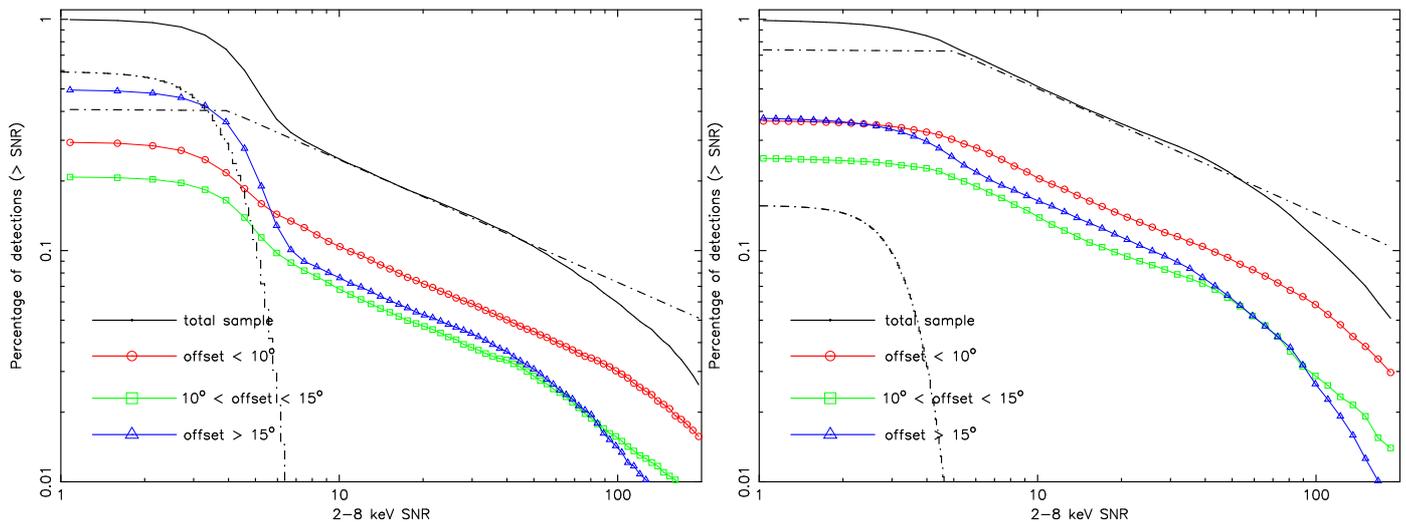

   \centering
   \resizebox{\hsize}{!}{ \rotatebox[]{-90}{\includegraphics[width=\textwidth]{verrecchia_fig5.ps}}

   \rotatebox[]{-90}{\includegraphics[width=\textwidth ]{verrecchia_fig6.ps}}
}
\vspace{-1.2cm}
     \caption{The cumulative 2--8\k\ SNR distributions of all the detections without any SNR cut for the first data processing run (left panel) and for the most complete one. The upper black thick curves are the total samples, the dash-dotted curves are the components of the fitted models (Gaussian + Broken power law, with parameters mean\,=\,3.9, $\sigma$\,=\,1.2, slopes $\sim$\,-0.53 and $\sim$\,-0.01 and break at $\sim$\,4.0 for the first processing run and mean\,=\,4.0, $\sigma$\,=\,1.2, same slopes and break at $\sim$\,5.8), while the remaining three lower curves are the distributions for different off-axis angle intervals (the circled one corresponds to off-axis angles $<$10\degr, the squared to 10--15\,$\degr$ and the triangled to $>$15$\degr$).
		  }
	   \label{Fig1d}
    \end{figure*}
%
%

 In order to distinguish those fields where the contribution to the background from bright sources is dominant, we considered the Total Count Rate (hereafter TCR) of an observation. TCR is defined as the total number of events divided by the net exposure time.
 Using this parameter we selected observations of low background fields (TCR\,$<$\,90\,$\ctss$ and TCR\,$<$\,130\,$\ctss$ in the first and second data processing run respectively\footnote{In the second processing run sources have different net exposures, so we built this parameter using the shortest common exposure time among them. We verified these ranges of values using the ``root mean squares'' of significance images discussed below in this section, as an alternative test of a field background.}), such as fields at high galactic latitude or those not crowded, and the high background ones at low latitude (TCR\,$>$\,90, 130\,$\ctss$).\\

The criterion to accept the detection of a point source is usually
different between known and new sources. 
The acceptance threshold of a known source at a certain position 
is defined as the level of 3$\sigma$ excess above the
statistical noise induced at that position by all contributions mentioned above (where $\sigma$
 is the statistical standard deviation). This corresponds, for a gaussian noise distribution, 
to a chance probability of less than 0.14\%.
 Since in a coded aperture camera the flux for every position in the sky is reconstructed from photons on
large portions of the detector, the noise distribution is still
Gaussian.

In the case of a previously uncatalogued X--ray source, its position is not known 
so one needs to take into account the number of trials.
 This number is firstly determined from the number of
independent pixels in the FOV, 255\,$\times$\,255 (i.e., the WFC number of mask elements). 
 If we require less than 1 false peak in 100 images, the threshold needs to be 5.3$\sigma$.

In our first procedure we implemented a very large reference catalogue, which produces in almost 
all WFC FOVs a large number of ``potentially'' visible known sources (see section 2.3). 
This increases proportionally the probability for false detections (1000 potential sources in the FOV
imply about 2 sources coincident by chance with 3$\sigma$ excesses).
This is not completely true however for GC fields, where the detectors non-linearity requires a fine tuning of the
detection acceptance threshold.

In order to take into account all the factors and evaluate the required significance threshold,
we built the cumulative distribution of all the 2--8\k\ detection SNRs (Fig. \ref{Fig1d}) in both data processing runs. This plot is roughly similar to a ``logN-logS'' distribution.

 We fit the data to a broken power law plus a cumulative gaussian model (BPL + CG) and in Fig. \ref{Fig1d} (left panel) we show both the components resulting from a fit to data from the first processing run stage (CG has a mean of 3.9 and $\sigma$\,$=$\,1.2, while the BPL has slopes of $\sim$\,-0.53, $\sim$\,-0.01 and a break at SNR\,$\sim$\,4.0). A BPL component was used to distinguish better the statistical noise component, represented by the CG component, from the true detections, represented by the BPL component. The BPL takes into account the instrumental sensitivity limits, the break roughly indicating the significance above which we can expect ``true'' detections following a power law distribution.

 The CG component, describing the background noise distribution (both random and systematic), has parameters which imply a sample of $\sim$4\% of spurious detections at SNR\,$\geq$\,6.3. We visually checked each marginal detection (with SNR$<$9, a distance from the reference position greater than 5\arcmin\ and at FOV borders).
 The percentage of discarded detections is $\sim$\,2.8\%, leaving about 1.2\% of residual noise detections in the first processing run sample.

In Fig. \ref{Fig1d} we also show the cumulative distributions for detections in three off-axis angle intervals, showing that the highest contribution to the CG noise component comes from detections at off-axis angles 
larger than 15\degr\ ($\sim$\,53\% of all detections in the gaussian component). These plots confirm the results of the visual inspection. Selecting detections with SNR greater than 6 from the first processing run, we obtained \dtnumproc\ detections of \fstnumproc\ distinct sources.

In the right panel of Fig. \ref{Fig1d} we show the 2--8\k\ SNR cumulative curves for the second processing run. These curves show a much lower CG component in all curves, evident only for off-axis angles larger than 15\degr. It was very difficult to estimate this component from the total curve. However the estimated CG parameters (mean\,$\sim$\,3.6 and $\sigma$\,$\sim$\,1.2) ensure a contamination from spurious detections ($\sim$\,1\% at SNR above 5.9) slightly larger than that obtained fitting the same component on the curve for off-axis angles larger than 15\degr. 
 This lower CG component however confirms the effect of the reference catalogue on the first processing run.

 In both processing runs the fitting process to the cumulative curves is not perfect. This is mainly due to the superposition of different noise components coming from different pointings (i.e., mainly from high/low background fields), but also to differences of all the other parameters which may influence the sensitivity among observations (such as different exposure times). Therefore, we first checked the SNR distributions for the low TCR sub-samples for both processing stages.
We estimated a gaussian component (mean\,$\sim$\,3.0 and $\sigma$\,$\sim$\,1.0) for the second stage which permits to establish a 1\% contamination threshold in this curve at SNR\,=\,4.8, while this component for the first processing run curve is comparable to that fitted on the total sample (threshold at 5.9).

 As a final verification of the acceptability of low SNR (below SNR\,=\,7) detections in low TCR fields, we computed specific 3\,$\sigma$ confidence level thresholds for each field estimating their background noise standard deviation\footnote{We evaluated these thresholds on significance images (the ratio of the sky image to the poisson noise map from the first iteration of IROS), either calculating the ``root mean square'' of a complete image and of sub-images (85$\times$85\,pixels) around each source, or estimating directly the gaussian background noise standard deviations. We corrected the 3\,$\sigma$ levels for the number of potential sources in each FOV.}, and checked visually detections above these thresholds.
 Most of the low SNR detections in low background fields were discarded with this check, up to ``single field'' confidence levels of $\sim$7$\sigma$.

   \begin{figure}[!ht]
   \centering
   \resizebox{\hsize}{!}{\rotatebox[]{-90}{\includegraphics[width=\textwidth ]{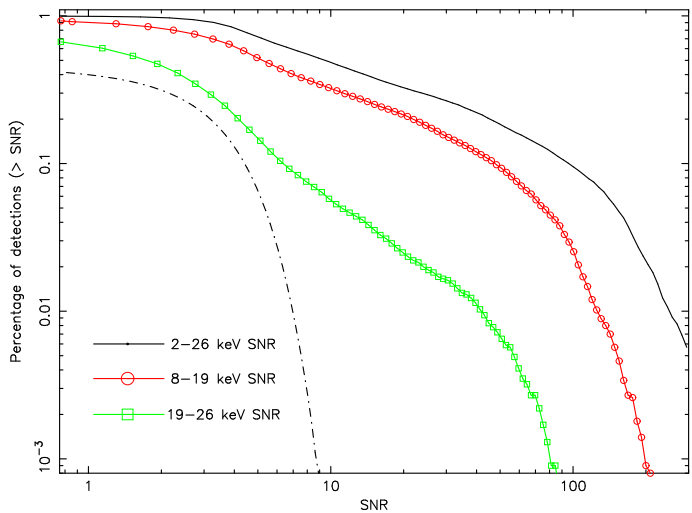}
  }}
 \vspace{-1.3cm}
   \caption{The cumulative distributions of SNR in three larger/harder bands (2--26, 8--19, 19--26\k) for detections of the second processing run. The gaussian component of the model fitted to the 8--19\k\ curve is shown.
		  }
	   \label{Fig1db}
    \end{figure}

We checked also results in harder bands from both processing runs, creating the same cumulative significance distributions (in Fig. \ref{Fig1db} those relative to the second processing run). In particular for the 8--19\k\ band we evaluated the CG component parameters (mean\,=\,2.0 and $\sigma$\,=\,2.7) using the same fit model, estimating a contribution of $\sim$\,2.5\% of noise detections above SNR\,=\,7.

The catalogue has been obtained choosing detections from the second processing run having SNR\,$>$\,6 if detected in the high TCR fields, or SNR\,$>$5 if in the low ones.
 Finally we added to the catalogue a small sample of verified sources from the first processing run and not detected in the second one together with a sample of detections from harder bands. 
The complete sample of detections includes \dtnum\ entries of \fstnum\ distinct sources.

\section{The Catalogue}

   The number of sources for the main classes of the \fstnum\ sources included in the catalogue are shown in Table \ref{Classes}. The positions and fluxes of the most significant detection of each source are reported in Table 3 (also available on-line at http://www.asdc.asi.it/wfccat/). At the end of the table a group of 11 rows includes sources detected in the first processing run only. 

   \begin{figure}[!ht]
   \centering
   \resizebox{\hsize}{!}{\rotatebox[]{-90}{\includegraphics[width=\textwidth]{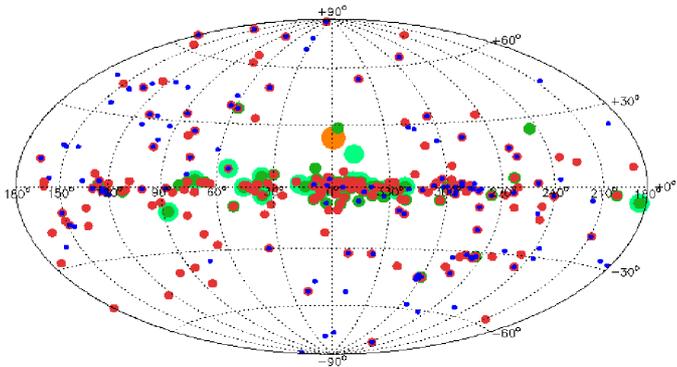}}}
\vspace{-2.4cm}
   \caption{The Aitoff projection of the \BWFC\ detections: \dtnum\ detections of \fstnum\ distinct X--ray sources in all the observations processed (symbol size is proportional to intensity)
 }
              \label{Fig3b}%
    \end{figure}

\begin{figure}[!ht]
\hspace{8.8cm}
\vspace{4.0cm}
\includegraphics{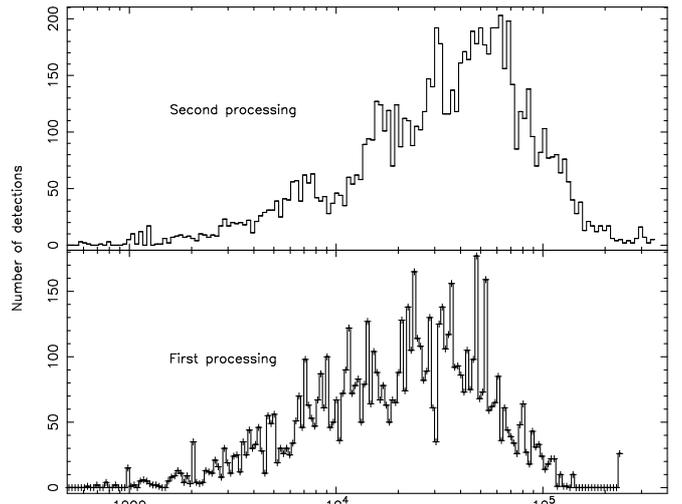}
\vspace{3.0cm}
\caption[t]{The exposure distributions of the \BWFC\ detections in all data processed. 
}
              \label{Fig14}%
\end{figure}

Table 3 gives as first column the \BWFC\ designation (i.e., ``SAXWFC''), followed by right ascension and declination (all celestial coordinates are for epoch J2000).
 In the following column the positional error is presented, which was calculated by a quadratic sum of the 1.3\arcmin\ systematic uncertainty (see below) and the mean of the 99\% confidence level errors in X and Y. Then are presented the source net exposure (in {\rm ks}), the 2--10\,\k\ flux and error (in \ergcms) for the maximum SNR detection, followed by the detection SNR in 2--8 and 8--19\,\k\ bands. The number of detections and the total exposure of each source are in the ninth and tenth column. The mean flux and its error (in \ergcms), the reduced chi-square and the minimum and maximum flux among all the detections of a source are in columns from the eleventh to the fourteenth.
The last two columns contain the source identification and the classification resulting from comparison with public databases.\\

Some statistical properties of the catalogue are hereafter described.
 The Aitoff projection in galactic coordinates of the full sample is shown in Fig. \ref{Fig3b}. The observation exposure distributions (Fig. \ref{Fig14}) have a peak at $\sim$\,30\,{\rm ks} and $\sim$\,50\,{\rm ks} for the first and second processing stages respectively, while only a few long high galactic latitude and GC pointings have exposures up to 350\,{\rm ks} in the second processing run. The total exposure is $\sim$\,67 Ms for the first processing run while the second one has a minimum total exposure of $\sim$\,102 Ms (choosing the minimum source exposure among those in a field, for each field). The catalogue mostly includes sources detected in the second processing stage.

   \begin{figure}[!ht]
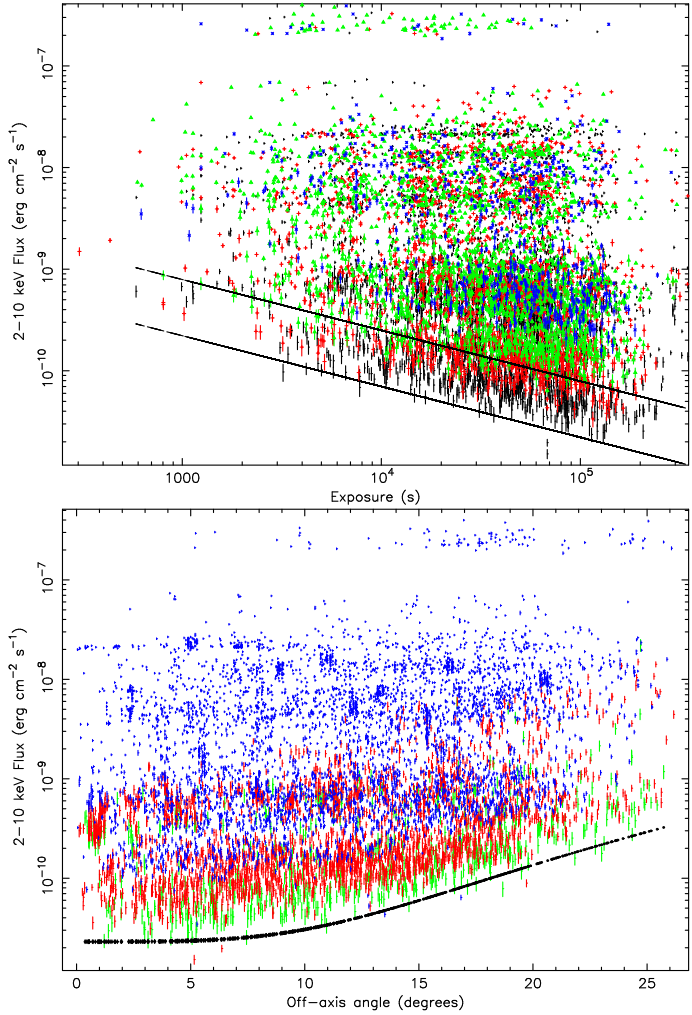

   \centering
   \resizebox{\hsize}{!}{\rotatebox[]{-90}{\includegraphics[width=\textwidth]{verrecchia_fig10.ps}
   \includegraphics[width=\textwidth]{verrecchia_fig11.ps}
}}
   \caption{The 2--10\,\k\ flux as a function of the exposure (top panel) of the final sample of detections.
 The dashed lines on the bottom part of the plot show the limiting sensitivity model (proportional to the exposure square root) for fields out of the Galactic Plane (the lowest) and the GC ones. In the bottom panel the same flux vs the off-axis angle.
 The dashed curve on the bottom part of the plot shows a model of the limiting sensitivity as a function of the off-axis angle.
   }
              \label{Fig17}
    \end{figure}
%

 The 2--10\,\k\ flux vs the exposures for all the detections in the sample in Fig. \ref{Fig17} (top panel) shows up the flux limit reached ($\sim$\,3\,$\times$\,10$^{-11}$~\ergcms) and the limiting sensitivity of the catalogue as a function of the exposure which follows the usual relation $F_x^{limit.}(2-10\,keV)\simeq c/T^{1/2}$, where the constant $c$ is
\nonumber
\be
 \mbox{ c}\simeq\left\{ \ba 7\times10^{-9}\,\,\ergcm{\rm s^{-1/2}} &\mbox{for $|b|>30^{\circ}$} \\
 3\times10^{-8}\,\,\ergcm{\rm s^{-1/2}} &\mbox{for GC region} \\
\ea \right .
\ee

\noindent These numbers are consistent with the predictions in \citealt*{Jag}. It should be noted that for GC fields the sensiti-\linebreak vity reaches a limit at about 100\,{\rm ks} of net exposure due to the above mentioned imperfect imaging.
 In the bottom panel of the same figure the 2--10\,\k\ flux as a function of the off-axis-angle is given. An empirical model for the WFC limiting sensitivity, $F_{x}^{limit.}$\,(2--10\,keV)\,$\simeq$\,(9.5\,$\times10^{-16} \times$\,{\em R}$^{3.9} + $2.3\,$\times10^{-11}$)\,\ergcms\,deg$^{-1}$, where {\em R} is the off-axis angle, has been evaluated for high galactic latitude fields.\\

   \begin{figure}[!ht]
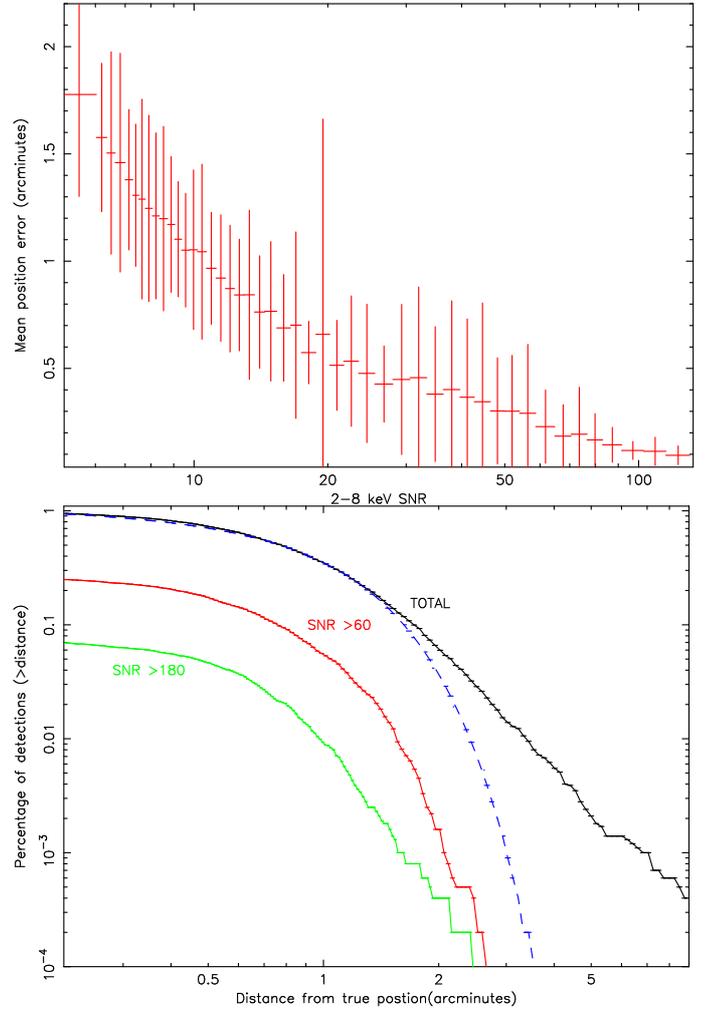

   \centering
   \resizebox{\hsize}{!}{\rotatebox[]{-90}{\includegraphics[width=\textwidth]{verrecchia_fig12.ps}\includegraphics[width=\textwidth]{verrecchia_fig13.ps}
}}
   \caption{Positional 99\% error radius (in arcmin) as a function of the SNR (top panel), where each data point represents the average over 150 source positions and the standard deviation with respect to that average. In the bottom panel the cumulative distribution of the distances between measured positions and those in the reference catalogue (in arcminutes) for all the detections (upper curve) and for detections with SNR\,$>$\,60 and SNR\,$>$\,180. The radii including 99\% of the detections are $\sim$\,3.3\arcmin, 1.87\arcmin\ and 1.78\arcmin\ respectively. The dashed curve is the cumulative gaussian component of the fit model applied to the upper curve, which represent the expected model for bright sources.
}
              \label{Fig20}%
    \end{figure}

%
\begin{figure}[!hb]
\centering
   \resizebox{\hsize}{!}{\rotatebox[]{-90}{
   \includegraphics[width=\textwidth]{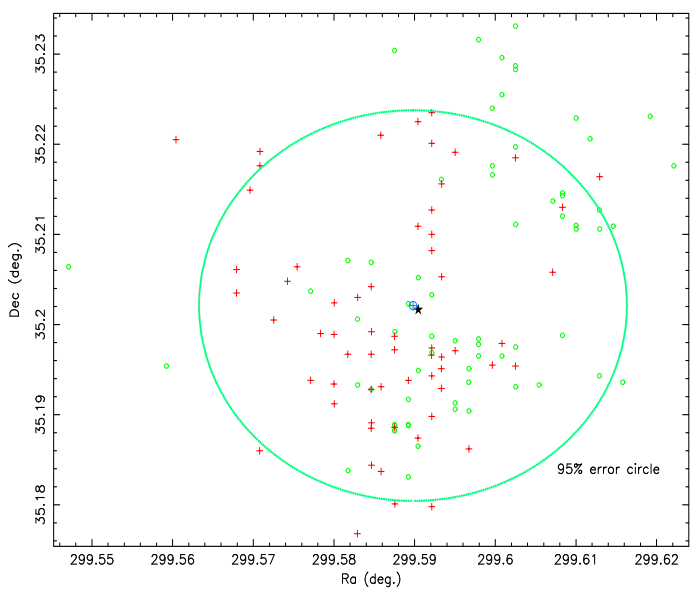}
}}
\vspace{-1.0cm}
     \caption[t]{
The positions of all Cyg X-1 detections in equatorial coordinates. Crosses correspond to WFC unit 1, circles to unit 2, while the star corresponds to the Cyg X-1 reference catalogue position, R.A.= 299.5904, Dec.= 35.2016 and the encircled cross to the best fit position, R.A.= 299.5898, Dec.= 35.2021. The thick circle of radius $\sim$\,1.6\arcmin\ represents the 95\,\% confidence level centered on the gaussian model best fit position.
 }
              \label{Fig4}%
\end{figure}
   \begin{figure}[!ht]
   \centering
   \resizebox{\hsize}{!}{\rotatebox[]{-90}{\includegraphics[width=\textwidth]{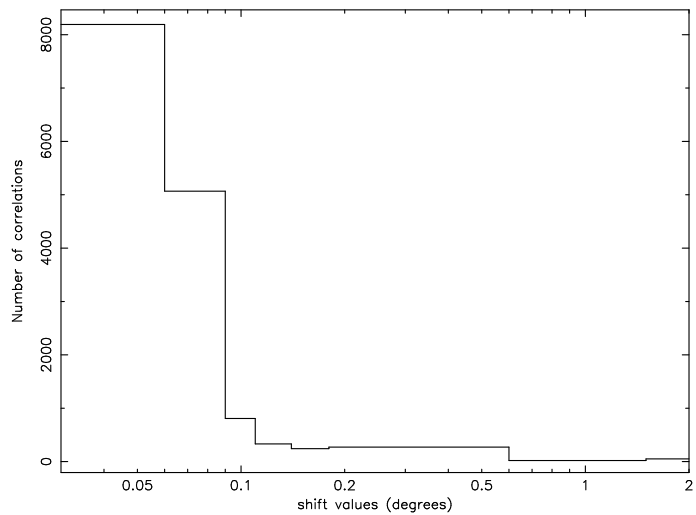}
}}
\vspace{-1.3cm}
   \caption{Spurious matches in the cross-correlation of the \BWFC\ catalogue with the reference catalogue with a radius of 5\arcmin\ for ``shift'' values from 0$^{\circ}$ to 2\,$^{\circ}$. }
              \label{Fig20b}%
    \end{figure}

The location accuracy reached is shown in Fig. \ref{Fig20}. Source location accuracy is known to depend on source SNR (\citealt*{Zand92}), so we plot the binned detection mean 99\% position errors as a function of the SNR (top panel). In each bin we calculate the error average value and its standard deviation over detections having similar SNRs. No systematic error appears at high SNRs. We show in the bottom panel the cumulative distributions of distances between measured and catalogued positions, for all the detections and for those with SNR\,$>$\,60 and SNR\,$>$\,180. Distances were obtained cross-correlating our list with the reference catalogue using a radius of 5\arcmin. We computed then radii including 99\% of the detections for all the distributions obtaining 3.30\arcmin, 1.87\arcmin\ and 1.78\arcmin\ for the total sample and the two sub-samples at increasing SNR thresholds respectively. 
We fitted these curves with a power law plus cumulative gaussian model to evaluate the 99\%\ confidence level error radius on the gaussian component, which is the expected distribution for bright ``good'' detections. The three curves illustrate how at higher SNR thresholds the distributions are better represented by the gaussian component. The radii for the three gaussian components are 1.78\arcmin, 1.61\arcmin\ and 1.33\arcmin\ respectively. This last value is consistent with what previously reported for bright sources (\citealt*{Add}, \citealt*{Hei}). This minimum location accuracy value has been added in quadrature as a systematic error to the mean 99\% position error for each detection. The total position error computed for the maximum SNR detection of each source is reported in Table 3.

As an example of the location accuracy for a bright source we show in Fig. \ref{Fig4} the positions of all Cyg X-1 detections in equatorial coordinates. We fitted the distribution of Cyg X-1 positions in both coordinates with a gaussian model obtaining a best fit mean position, R.A.= 299.5898, Dec.= 35.2021 (J2000) with a $\sim$\,1.6\arcmin\ 95\% confidence level radius.

 The chance coincidence probability of the correlation between WFC detections and the reference catalogue positions used in the source post-processing identification, can be evaluated by shifting the detection positions by various fixed amounts (between -10$^{\circ}$ and 10\,$^{\circ}$) and executing again the cross-correlation with the same radius (see Fig. \ref{Fig20b}).
The number of spurious matches for shift values greater than 0.16$^{\circ}$ is below $\sim$\,3\%.

\begin{figure*}[htp]
   \centering
   \resizebox{\hsize}{!}{\includegraphics[width=1.4\textwidth]{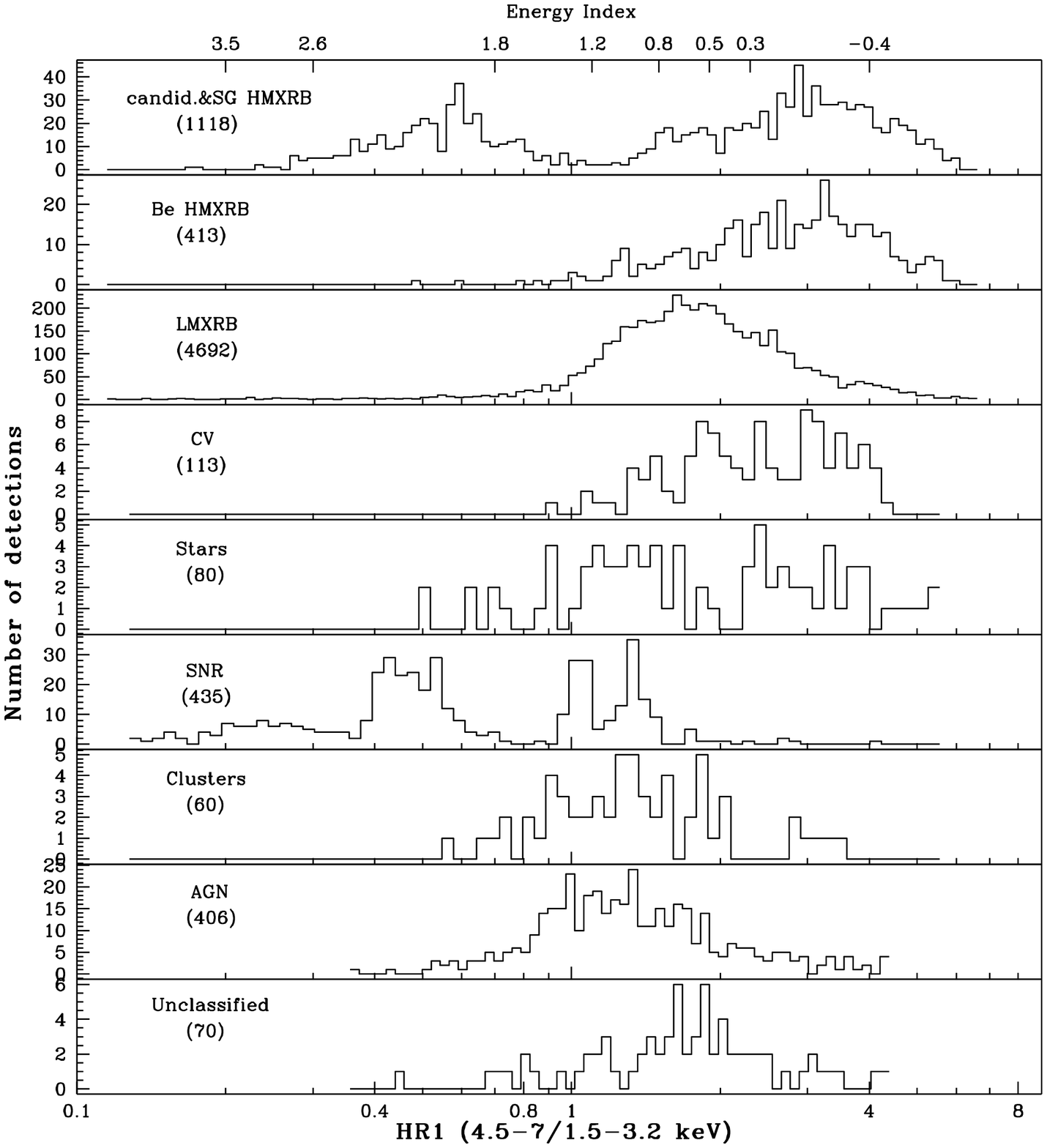}
 \hspace{0.7cm}
\vspace{-2cm}
\includegraphics[width=1.4\textwidth]{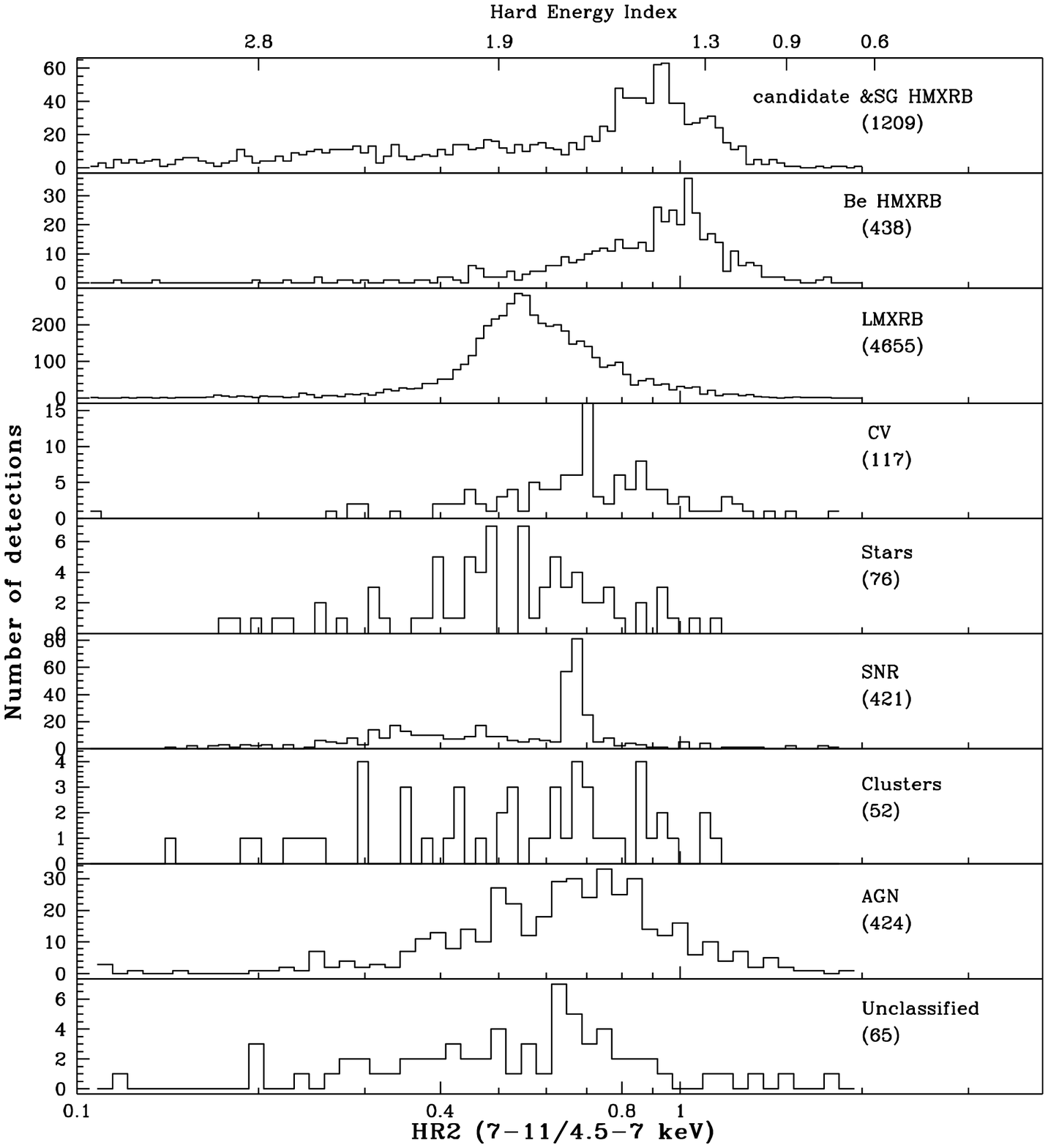}
 }
     \caption{
The Hardness Ratio 1 (1.7--3.2/4.6--7.1\,\k\ bands; left panel) distributions for each source class included in the catalogue. On the right the same per classes distributions of Hardness Ratio 2 (4.6--7.1/7.1--10.9\,\k). Two corresponding energy indexes, evaluated assuming galactic absorption, are shown in the top scales of both panels. Numbers below each class label, in both panels, are the detections having non-null values of count rates and of energy indexes.
		}
	 \label{Fig18-19}%
   \end{figure*}

We calculated two hardness ratios (HR1, HR2) using the 1.7--3.2, 4.6--7.1 and 7.1--10.9\,\k\ bands for two sub-samples of \colsam\ and \colcolsam\ detections respectively, having non-zero values of count rates in these bands. The HR1 distribution for each individual source class included in the catalogue (see Table \ref{Classes}) is shown in Fig. \ref{Fig18-19}.
These histograms include HR1 values from all possible sources states, such as hard/soft states for LMXRBs and HMXRBs, all overlapped. In particular for Supergiant plus unclassified HMXRBs HR1 covers a very broad range of values with two peaks at $\sim$\,0.6 and $\sim$\,3.

The HR2 distributions for each source class are shown in the left panel of Fig. \ref{Fig18-19}. In this case the plot for Supergiant HMXRBs class shows a prevailing harder peak at $\sim$\,1, where the Be HMXRBs distribution peaks too.
 Two energy indexes were evaluated for both HRs assuming galactic absorption values for all sources. They are shown in the two panels of the Fig. as top scales. Their values range between -0.6 and 3.5 for that evaluated from HR1 and between 0.5 and 3.0 for the harder one.

\section{Conclusions}

 We presented the complete post-mission catalogue of WFC X--ray sources detectable in complete observations, which includes \fstnum\ distinct objects.
 As shown in Table \ref{Classes}, these sources are mainly galactic (176 sources plus three candidate galactic transient sources, i.e. 71\% of the total), while the smaller extra-galactic sample, 62 sources (25\%), is mainly composed of AGN, 11 of which are BL Lacs (4.3\%). The galactic sources are mostly LMXRBs (48\%) while the HMXRBs (27\%) include 13 classified Supergiant systems (7.3\%) and 23 Be HMXRBs (13\%). The 14 unclassified sources in Table 3, include  
 two known sources, SAX J1428.6-5422 and SAX J1805.5-2031, and a new transient X--ray source, for which the WFC detections give the best localization so far. These are candidate transient galactic sources detected in outburst. Also included is a more uncertain candidate X--ray binary source (RX J0121.4-7258) and a possible BL Lac object (1RXS J205528.2-002123).

 Most of the catalogue sources (83\%) have a counterpart in the softer RASS catalogue (within 3\arcmin\ radius). Moreover large fractions of sources have significant (SNR\,$\geq$\,7) detections in the 2-5.5, 2-19 and 8-19\k\ bands (219, 197 and 133 sources respectively), which have slightly lower percentages (71\%, 70\% and 64\% respectively) of counterparts in the RASS.
 On the other hand the correlation with the harder INTEGRAL/IBIS Soft $\gamma$-ray catalogues (Bird et al. 2004, 2006 and 2007) give increasing percentages proportionally to the increasing INTEGRAL total exposure and sky coverage (29\% with the first, 40\% with the second and 53\% with the most recent one, using the same 3\arcmin\ radius). If we consider a sample of harder WFC sources (35\%), detected (SNR$\geq$10) in the 11-26\k\ band, 78\% resulted to have a counterpart in the last IBIS catalogue.  
The WFC catalogue thus includes sources detected up to 19\k\ with still a good sensitivity (sources detected in this band have a 2--10\k\ flux greater than about 3 mCrabs).

The cross-correlation with the RXTE ``All Sky Mo\-ni\-tor'' source catalogue (available on-line\footnote{http://\-xte.mit.edu/asmlc/asmcat.html}) finally gives the highest number of results (75\%) among catalogues in hard X--ray bands. This result confirms the importance of the WFC source catalogue as a complementary reference catalogue for current and future hard X-ray missions, particularly in the 2--10\,\k\ band.

\begin{acknowledgements}

 The ASDC is supported by the Italian Space Agency (ASI). Part of this work benefitted of ASI grant I/024/05. FV, PG and PS thank S. Piranomonte for the help to the revision of source identifications and classes detected in the first data processing run. Authors thank all people of the WFC team at SRON, particularly G. Wiersma, for his collaboration and kindness.

\end{acknowledgements}

\newpage
\authorrunning{}
\titlerunning{}
\topmargin 2.cm
\thispagestyle{empty}

\markright{}
\vspace{3.0cm}
\vskip 18.0cm
\centering \hfill \hfill \hfill \hfill \large \BWFC\ Catalogue
\vspace{3.0cm}

\centering \hfill \hfill \hfill \hfill \large {Supplementary Table (Table 3)}

\thispagestyle{empty}

\def\ergs{\rm erg\,s^{-1}}
\def\ergcs{\rm erg\,cm$^{-2}$\,s$^{-1}$}
\textwidth 20.0cm
\textheight 30.0cm
\topmargin 4cm
\oddsidemargin -1cm


\thispagestyle{empty}
\begin{landscape}

\markright{}
\authorrunning{}
\titlerunning{}

\begin{table*}[tbh]
\setcounter{table}{2}

\scriptsize

\begin{minipage}{250mm}
\caption{ BeppoSAX WFC X-ray Source Catalogue}

\begin{tabular}{rccccrrrrrrlllll}
\hline
\hline
 WFC Name & RA(2000) & DEC(2000) & $^{a}$ERR$_{RAD}$ & EXPO &
   $^{b}$Flux$_{2-10}$ & $^{c}$SNR1 & $^{d}$SNR2 & N$_{det}$ &
   EXPO$_{total}$ & $^{e}<$FLUX$>_{2-10}$ & $\chi^{2}_{flux}$  &
   $^{f}$Flux$_{MAX}$  & $^{f}$Flux$_{MIN }$ & Name & $^{g}$Class   \\
   & hh mm ss & o  '  "   & arcmin & ks~~~ & &  &  &   & ks ~~~~~ &  & &  & &  &
  \\
\hline
SAXWFC J0025.5+6407.3 & 00 25 28.7 &  64 07 18.8 & 1.3 &    74.0 &    17.6 $\pm$ 0.8 &   24.9 &    1.5 & 105 & 6358.6 & 17.1 $\pm$  2.9 & 3.5 & 28.0 & 11.0 & Tycho & SNR shell   \\ 
SAXWFC J0028.7+5917.2 & 00 28 44.7 &  59 17 11.7 & 3.5 &    45.3 &     4.7 $\pm$ 0.6 &    7.8 &    2.8 & 9 & 760.2 &  5.1 $\pm$  1.3 & 1.6 &  8.6 &  3.9 & V709 Cas & CV    \\ 
SAXWFC J0035.9+5949.7 & 00 35 52.2 &  59 49 42.2 & 1.5 &   101.2 &     8.2 $\pm$ 0.7 &   12.7 &    3.1 & 10 & 809.3 &  6.6 $\pm$  1.9 & 4.5 & 11.0 &  4.5 & 1ES 0033+595 & AGN, BL Lac   \\ 
SAXWFC J0037.0+6121.0 & 00 37 01.7 &  61 21 02.8 & 1.4 &    46.4 &    12.0 $\pm$ 0.7 &   16.9 &    5.4 & 6 & 357.6 &  9.6 $\pm$  7.6 & 27.9 & 27.0 &  6.0 & IGR J00370+6122 & HMXRB    \\ 
SAXWFC J0051.8-7217.5 & 00 51 45.0 & -72 17 31.2 & 1.5 &    33.9 &     9.3 $\pm$ 0.6 &   15.8 &    6.4 & 1 & -- & --- & -- & -- & -- & SMC X-3 & HMXRB, XP   \\ 
SAXWFC J0052.3-7319.3 & 00 52 20.1 & -73 19 17.0 & 1.3 &   169.5 &    15.6 $\pm$ 0.6 &   26.8 &   10.7 & 12 & 571.1 & 17.0 $\pm$  7.3 & 23.8 & 35.0 &  6.2 & RX J0052-7319 & Be HMXRB, XP, T  \\ 
SAXWFC J0053.2-7438.8 & 00 53 09.5 & -74 38 50.2 & 1.6 &   109.8 &     6.3 $\pm$ 0.6 &   11.1 &    5.4 & 2 & 174.7 &  6.6 $\pm$  0.5 & 0.8 &  7.1 &  6.3 & CF Tuc & RSCVn  \\ 
SAXWFC J0054.3-7340.2 & 00 54 19.7 & -73 40 12.7 & 2.1 &    76.6 &    13.1 $\pm$ 2.0 &    5.7 &    4.0 & 2 & 240.9 & 10.2 $\pm$  3.0 & 3.0 & 13.0 &  8.8 & SMC X-2 & Be HMXRB, XP, T  \\ 
SAXWFC J0056.7+6043.4 & 00 56 44.9 &  60 43 26.7 & 1.3 &   248.0 &    16.3 $\pm$ 0.6 &   32.3 &   11.1 & 72 & 4321.3 & 14.2 $\pm$  3.6 & 9.2 & 24.0 &  7.7 & Gamma Cas & Be HMXRB  \\ 
SAXWFC J0111.3-7317.2 & 01 11 18.1 & -73 17 13.2 & 1.3 &   140.0 &    57.2 $\pm$ 1.2 &  124.2 &   59.3 & 17 & 1083.6 & 17.4 $\pm$ 20.5 & 300.6 & 62.0 &  4.8 & XTE J0111-7317 & Be HMXRB, XP, T  \\ 
SAXWFC J0117.0-7326.4 & 01 17 00.5 & -73 26 22.5 & 1.3 &   103.8 &    81.5 $\pm$ 1.7 &  114.3 &   61.0 & 158 & 11287.8 & 39.6 $\pm$ 26.2 & 282.6 & 127.0 &  5.5 & SMC X-1 & SG HMXRB, XP   \\ 
SAXWFC J0118.0+6517.7 & 01 18 02.3 &  65 17 42.7 & 1.3 &    39.5 &    28.3 $\pm$ 0.7 &   57.7 &   44.9 & 48 & 2978.6 & 14.2 $\pm$  7.6 & 46.9 & 39.0 &  7.3 & 2S 0114+65 & SG HMXRB, XP   \\ 
SAXWFC J0118.5+6343.8 & 01 18 29.8 &  63 43 47.2 & 1.3 &   108.3 &   163.5 $\pm$ 3.6 &   79.2 &   56.1 & 10 & 535.3 & 31.9 $\pm$ 126.2 & 450.1 & 420.0 & 10.0 & 4U 0115+634 & Be HMXRB, XP, T  \\ 

SAXWFC J0121.7-7257.6 & 01 21 41.0 & -72 57 33.4 & 1.4 &    64.9 &    16.6 $\pm$ 0.7 &   24.1 &   19.7 & 2 & 157.7 & 17.5 $\pm$  2.7 & 6.6 & 20.0 & 17.0 & RX J0121.4-7258 & unclassified    \\ 
SAXWFC J0124.0-3504.6 & 01 23 57.0 & -35 04 36.1 & 2.0 &    69.1 &     6.2 $\pm$ 0.6 &    9.5 &    4.0 & 3 & 141.8 &  7.1 $\pm$  6.5 & 12.3 & 18.0 &  6.2 & NGC 526A & AGN, Seyfert 1  \\ 
SAXWFC J0146.5+6145.3 & 01 46 32.2 &  61 45 15.4 & 1.5 &   259.5 &     8.3 $\pm$ 0.6 &   13.5 &    0.8 & 31 & 2368.4 &  7.7 $\pm$  2.0 & 3.2 & 16.0 &  5.5 & 4U 0142+614 & AXP   \\ 
SAXWFC J0147.0+6120.6 & 01 47 01.2 &  61 20 34.0 & 1.5 &    81.9 &     9.8 $\pm$ 0.7 &   14.0 &    6.5 & 23 & 1875.3 &  7.7 $\pm$  5.2 & 8.9 & 31.0 &  4.2 & RX J0146.9+6121 & Be HMXRB, XP  \\ 
SAXWFC J0209.5+5226.2 & 02 09 31.8 &  52 26 13.2 & 2.3 &    59.0 &    15.9 $\pm$ 2.7 &    5.1 &    0.3 & 1 & -- & --- & -- & -- & -- & 1H 0203+513 & AGN, Seyfert 1  \\ 
SAXWFC J0245.0+6225.7 & 02 44 57.5 &  62 25 42.6 & 1.7 &    38.0 &     6.5 $\pm$ 0.7 &    9.7 &    3.6 & 7 & 535.6 &  5.1 $\pm$  0.7 & 1.2 &  6.5 &  4.2 & QSO B0241+62 & AGN, Seyfert 1  \\ 
SAXWFC J0254.3+4133.3 & 02 54 20.7 &  41 33 17.2 & 1.9 &   103.4 &     5.9 $\pm$ 0.9 &    6.6 &    1.5 & 3 & 333.1 &  5.2 $\pm$  0.6 & 0.5 &  5.9 &  4.7 & 2A 0251+413 & ClG    \\ 
SAXWFC J0308.2+4057.1 & 03 08 11.9 &  40 57 08.6 & 1.3 &    20.9 &    68.9 $\pm$ 2.4 &   31.0 &    2.8 & 6 & 241.3 & 11.5 $\pm$ 23.7 & 165.3 & 69.0 &  4.1 & Algol & Algol star type   \\ 
SAXWFC J0311.3-2046.2 & 03 11 20.5 & -20 46 09.1 & 2.1 &    14.4 &    12.7 $\pm$ 2.0 &    6.0 &    0.0 & 1 & -- & --- & -- & -- & -- & HE 0309-2057 & AGN, Seyfert 1    \\ 

SAXWFC J0318.0-4414.7 & 03 18 00.0 & -44 14 44.5 & 2.1 &    82.7 &     5.6 $\pm$ 0.8 &    7.0 &    4.2 & 3 & 328.9 &  3.5 $\pm$  1.6 & 5.2 &  5.6 &  2.8 & ABELL 3112 & ClG  \\ 
SAXWFC J0319.8+4131.0 & 03 19 47.1 &  41 31 01.2 & 1.3 &    25.0 &    40.6 $\pm$ 1.6 &   29.5 &    6.0 & 40 & 1377.8 & 37.3 $\pm$  3.7 & 2.6 & 49.0 & 32.0 & NGC 1275 & AGN, RG  \\ 
SAXWFC J0326.6+2842.7 & 03 26 33.0 &  28 42 44.2 & 1.5 &    81.9 &     9.9 $\pm$ 0.6 &   15.5 &    3.8 & 2 & 157.6 &  9.3 $\pm$  1.5 & 3.1 &  9.9 &  7.9 & UX Ari & RSCVn  \\ 
SAXWFC J0331.3+4356.0 & 03 31 16.6 &  43 56 01.6 & 1.7 &    64.4 &     9.6 $\pm$ 0.9 &   10.2 &    4.5 & 3 & 224.5 & 11.7 $\pm$  6.0 & 11.0 & 22.0 &  9.6 & GK Per & CV, IP \\ 

SAXWFC J0336.8+0034.9$^{h}$ & 03 36 46.5 &  00 34 55.5 & 1.6 &    78.2 &     0.0 $\pm$ 0.0 &    0.0 &    0.0 & 3 & 251.7 & 12.9 $\pm$ 19.7 & 5.4 & 49.0 & 10.0 & V711 Tau & RSCVn  \\ 
SAXWFC J0355.4+3103.3 & 03 55 22.9 &  31 03 20.8 & 1.3 &   164.8 &    30.2 $\pm$ 0.9 &   38.8 &   13.1 & 25 & 1061.3 & 21.4 $\pm$  5.0 & 10.3 & 35.0 & 16.0 & 4U 0352+309 & Be HMXRB, XP  \\ 
SAXWFC J0413.4+1026.6 & 04 13 23.8 &  10 26 34.0 & 2.0 &    50.3 &     7.9 $\pm$ 1.1 &    6.7 &    0.0 & 1 & -- & --- & -- & -- & -- & ABELL 478 & ClG  \\ 
SAXWFC J0433.2+0521.1 & 04 33 09.3 &  05 21 06.1 & 2.1 &    50.1 &     8.2 $\pm$ 1.1 &    7.3 &    1.0 & 2 & 89.5 &  6.3 $\pm$  2.2 & 5.0 &  8.2 &  5.1 & 3C 120 & AGN, Seyfert 1  \\ 
SAXWFC J0452.0+4932.1 & 04 52 02.9 &  49 32 08.5 & 1.8 &   135.5 &     4.6 $\pm$ 0.5 &    8.1 &    0.0 & 5 & 365.5 &  4.4 $\pm$  1.1 & 3.2 &  6.3 &  3.2 & 1RXS J045205.0+493248 & AGN, Seyfert 1  \\ 
SAXWFC J0458.2-7516.4 & 04 58 14.0 & -75 16 21.3 & 1.8 &    70.5 &     8.6 $\pm$ 1.0 &    7.4 &    1.6 & 3 & 116.8 &  9.3 $\pm$  1.0 & 0.6 & 11.0 &  8.6 & 1RXS J045816.3-751608 & star KIII   \\ 
SAXWFC J0507.7+6738.3 & 05 07 44.5 &  67 38 16.4 & 1.7 &   208.0 &     3.9 $\pm$ 0.4 &    9.5 &    2.5 & 6 & 977.3 &  4.3 $\pm$  1.6 & 2.6 &  7.9 &  3.8 & 1ES0502+675 & AGN, BL Lac   \\ 
SAXWFC J0514.1-4002.3 & 05 14 07.3 & -40 02 19.3 & 1.3 &    46.6 &    34.9 $\pm$ 0.9 &   56.4 &   16.5 & 38 & 1567.5 & 11.0 $\pm$  7.6 & 55.2 & 35.0 &  4.5 & 4U 0513-40 & LMXRB, GlC, B  \\ 
SAXWFC J0520.4-7157.6 & 05 20 24.4 & -71 57 37.0 & 1.3 &   191.8 &    53.1 $\pm$ 1.1 &  118.5 &   13.0 & 269 & 16241.2 & 51.3 $\pm$ 10.0 & 15.1 & 102.0 & 27.0 & LMC X-2 & LMXRB    \\ 
SAXWFC J0528.8-6527.0 & 05 28 46.6 & -65 27 00.3 & 1.5 &     9.1 &    45.9 $\pm$ 3.2 &   13.4 &    2.6 & 6 & 303.2 &  9.4 $\pm$ 15.1 & 31.0 & 46.0 &  5.6 & AB Dor & star K1IV   \\ 
SAXWFC J0529.2-3248.3 & 05 29 13.6 & -32 48 18.3 & 1.8 &    82.3 &     7.3 $\pm$ 1.0 &    6.8 &    2.5 & 4 & 190.6 &  6.8 $\pm$  1.0 & 1.3 &  7.7 &  5.5 & TV Col & CV    \\ 
SAXWFC J0532.9-6622.0 & 05 32 54.5 & -66 22 00.1 & 1.3 &    61.5 &    16.8 $\pm$ 0.6 &   31.4 &   30.0 & 67 & 3706.5 & 15.4 $\pm$  7.9 & 31.9 & 49.0 &  3.6 & LMC X-4 & SG HMXRB, XP  \\ 
SAXWFC J0534.5+2201.2 & 05 34 30.9 &  22 01 09.4 & 1.3 &    23.5 &  2071.4 $\pm$ 41.5 &  751.4 &  155.0 & 152 & 4635.1 & 2113.8 $\pm$ 89.4 & 5.1 & 2571.0 & 1680.7 & Crab Nebula & SNR filled-center, XP   \\ 
SAXWFC J0535.3-0525.4 & 05 35 18.7 & -05 25 21.0 & 1.9 &    30.5 &     6.4 $\pm$ 1.1 &    5.9 &    0.0 & 1 & -- & --- & -- & -- & -- & AXS J053514-0523 & unclassified    \\ 
SAXWFC J0539.0-6405.2 & 05 38 57.5 & -64 05 15.0 & 1.3 &    78.3 &    66.8 $\pm$ 1.4 &  122.4 &    5.1 & 167 & 7781.3 & 36.8 $\pm$ 25.2 & 192.5 & 215.0 &  5.9 & LMC X-3 & HMXRB, BH  \\ 
SAXWFC J0539.6-6944.7 & 05 39 38.0 & -69 44 44.1 & 1.3 &    69.5 &    49.6 $\pm$ 1.1 &   99.7 &    5.8 & 209 & 11334.9 & 39.7 $\pm$  7.8 & 14.2 & 77.0 & 24.0 & LMC X-1 & SG HMXRB, BH  \\ 
SAXWFC J0540.1-6919.2 & 05 40 05.3 & -69 19 14.8 & 1.7 &    30.7 &     5.9 $\pm$ 0.8 &    7.8 &    2.9 & 6 & 720.7 &  4.1 $\pm$  1.0 & 2.1 &  5.9 &  3.3 & PSR J0540-6919 & XP    \\ 
SAXWFC J0542.8+6051.2 & 05 42 50.5 &  60 51 11.1 & 1.8 &   161.6 &     4.6 $\pm$ 0.6 &    7.9 &    3.4 & 4 & 611.7 &  4.3 $\pm$  0.4 & 0.5 &  4.6 &  3.8 & BY Cam & CV, AM Her  \\ 
SAXWFC J0550.6-3216.2 & 05 50 34.9 & -32 16 11.2 & 1.7 &    91.7 &     4.5 $\pm$ 0.5 &    9.0 &    1.6 & 2 & 172.2 &  4.7 $\pm$  0.8 & 1.2 &  5.6 &  4.5 & PKS 0548-322 & AGN, BL Lac   \\ 
SAXWFC J0552.1-0727.9$^{j}$ & 05 52 06.3 & -07 27 51.1 & 2.0 &    44.8 &     0.0 $\pm$ 0.0 &    0.0 &    0.0 & 3 & 302.4 & 13.2 $\pm$ 24.6 & 6.2 & 57.0 &  3.0 & NGC 2110 & AGN, Seyfert 2  \\ 
SAXWFC J0552.5+5928.9 & 05 52 28.2 &  59 28 53.0 & 2.0 &   135.5 &     4.2 $\pm$ 0.6 &    6.7 &    1.1 & 1 & -- & --- & -- & -- & -- & IRAS 05480+5927 & AGN, Seyfert 1  \\ 
SAXWFC J0554.9+4625.6 & 05 54 55.2 &  46 25 33.6 & 1.5 &   125.6 &     6.9 $\pm$ 0.5 &   14.7 &    5.2 & 7 & 377.2 &  6.9 $\pm$  1.8 & 3.6 & 11.0 &  5.7 & MCG+08-11-011 & AGN, Seyfert 1  \\ 
SAXWFC J0617.1+0908.9 & 06 17 07.4 &  09 08 51.3 & 1.3 &    80.2 &   129.9 $\pm$ 2.9 &  105.3 &   27.6 & 95 & 1908.9 & 83.1 $\pm$ 46.7 & 53.6 & 360.0 & 37.0 & 4U 0614+091 & LMXRB, B   \\ 
SAXWFC J0620.8+1327.0 & 06 20 49.9 &  13 27 01.0 & 2.3 &    97.8 &    43.0 $\pm$ 6.9 &    5.5 &    1.8 & 1 & 326.9 & --- & -- & -- & -- & WD 0618+134 & WD, DA \\ 
SAXWFC J0637.0-0522.3 & 06 36 57.8 & -05 22 16.3 & 1.7 &    42.3 &     9.7 $\pm$ 0.7 &   13.0 &    1.8 & 1 & -- & --- & -- & -- & -- & 1RXS J063656.7-052104 & unclassified    \\ 
SAXWFC J0714.3-3625.4 & 07 14 16.4 & -36 25 23.1 & 2.1 &    23.8 &     9.7 $\pm$ 1.3 &    7.1 &    1.6 & 1 & -- & --- & -- & -- & -- & 1RXS J071413.7-362533 & unclassified   \\ 
SAXWFC J0728.9-2606.4 & 07 28 51.6 & -26 06 24.8 & 1.5 &    62.8 &    17.0 $\pm$ 1.5 &   10.4 &    5.2 & 5 & 248.2 & 10.2 $\pm$  4.4 & 10.4 & 17.0 &  7.3 & 4U 0728-25 & Be HMXRB, XP  \\ 
SAXWFC J0743.2+2853.1 & 07 43 12.8 &  28 53 06.0 & 2.0 &    99.5 &     5.6 $\pm$ 0.8 &    6.8 &    0.0 & 1 & -- & --- & -- & -- & -- & Sig Gem & RSCVn  \\ 
SAXWFC J0747.5-1917.6 & 07 47 30.4 & -19 17 35.1 & 1.7 &    46.7 &     6.3 $\pm$ 0.7 &    9.5 &    0.0 & 4 & 294.5 &  6.2 $\pm$  0.4 & 0.2 &  6.7 &  5.8 & PKS 0745-19 & ClG  \\ 
\hline

\end{tabular}
\end{minipage}
\end{table*}

\newpage
\thispagestyle{empty}
\markright{}

\begin{table*}[tbh]
\setcounter{table}{2}

\scriptsize

\begin{minipage}{250mm}
\caption{ BeppoSAX WFC X-ray Source Catalogue}

\begin{tabular}{rccccrrrrrrlllll}
\hline
\hline
 WFC Name & RA(2000) & DEC(2000) & $^{a}$ERR$_{RAD}$ & EXPO &
   $^{b}$Flux$_{2-10}$ & $^{c}$SNR1 & $^{d}$SNR2 & N$_{det}$ &
   EXPO$_{total}$ & $^{e}<$FLUX$>_{2-10}$ & $\chi^{2}_{flux}$  &
   $^{f}$Flux$_{MAX}$  & $^{f}$Flux$_{MIN }$ & Name & Class   \\
   & hh mm ss & o  '  "   & arcmin & ks~~~ & &  &  &   & ks ~~~~~ &  & &  & &  &
  \\
\hline
SAXWFC J0748.5-6745.2 & 07 48 28.8 & -67 45 14.7 & 1.3 &    56.6 &    43.2 $\pm$ 1.2 &   49.1 &   11.8 & 180 & 9893.5 & 15.8 $\pm$  7.4 & 14.6 & 63.0 &  8.3 & EXO 0748-676 & LMXRB, B, E  \\ 
SAXWFC J0749.1-0551.3 & 07 49 06.7 & -05 51 16.2 & 2.2 &    42.5 &    51.7 $\pm$ 7.2 &    6.0 &    6.5 & 1 & -- & --- & -- & -- & -- & RX J0749.1-0549 & CV, AM Her  \\ 

SAXWFC J0812.5-3114.3 & 08 12 29.4 & -31 14 17.8 & 1.5 &    47.3 &    11.7 $\pm$ 0.6 &   19.1 &    5.8 & 10 & 330.2 & 12.1 $\pm$ 10.8 & 31.5 & 42.0 &  5.2 & RX J0812.4-3114 & Be HMXRB, XP, T  \\ 
SAXWFC J0823.2-4250.1$^{h}$ & 08 23 11.7 & -42 50 04.2 & 1.5 &    41.7 &    0.0 $\pm$ 0.0 &    0.0 &    0.0 & 28 & 1274.8 &  5.7 $\pm$  5.4 & 2.1 & 19.0 &  4.8 & 1ES 0821-426 & unclassified    \\ 
SAXWFC J0835.3-4510.8 & 08 35 19.3 & -45 10 50.5 & 1.6 &    42.5 &     6.7 $\pm$ 0.7 &   10.0 &    5.1 & 18 & 1086.5 &  7.3 $\pm$  1.2 & 1.2 & 11.0 &  5.5 & Vela Pulsar & SNR, XP, RP  \\ 
SAXWFC J0841.4+7054.8 & 08 41 21.7 &  70 54 48.2 & 1.9 &    60.5 &     6.4 $\pm$ 0.8 &    8.1 &    2.0 & 10 & 924.9 &  4.9 $\pm$  1.0 & 1.6 &  6.6 &  4.1 & 4C 71.07 & QSO  \\ 
SAXWFC J0902.1-4032.8 & 09 02 04.2 & -40 32 46.6 & 1.3 &   102.0 &   187.0 $\pm$ 3.7 &  221.0 &   86.5 & 129 & 4770.0 & 82.5 $\pm$ 76.4 & 483.7 & 350.0 & 14.0 & Vela X-1 & SG HMXRB, XP, T   \\ 
SAXWFC J0920.4-5512.4 & 09 20 26.1 & -55 12 24.1 & 1.3 &     6.0 &   174.3 $\pm$ 3.8 &  107.0 &   38.0 & 109 & 5273.6 & 14.4 $\pm$ 16.9 & 55.7 & 170.0 &  4.0 & 2S 0918-549 & LMXRB, B    \\ 
SAXWFC J0922.6-6317.6 & 09 22 38.4 & -63 17 35.1 & 1.5 &    44.9 &     8.4 $\pm$ 0.7 &   11.0 &    2.6 & 21 & 1473.7 &  6.5 $\pm$  2.4 & 8.2 & 12.0 &  3.6 & 2S 0921-63 & LMXRB, E  \\ 
SAXWFC J0923.7+2253.8 & 09 23 39.7 &  22 53 46.3 & 2.3 &    26.1 &     9.8 $\pm$ 1.2 &    7.4 &    0.0 & 1 & -- & --- & -- & -- & -- & RX J0923.7+2254 & AGN, Seyfert 1  \\ 
SAXWFC J0924.3-3142.4 & 09 24 17.7 & -31 42 24.8 & 1.6 &    74.7 &     9.7 $\pm$ 0.6 &   14.5 &    0.0 & 32 & 1456.1 & 12.3 $\pm$  4.3 & 6.7 & 26.0 &  8.5 & 1RXS J092418.0-314212 & unclassified    \\ 
SAXWFC J0934.1+6251.9 & 09 34 06.0 &  62 51 55.8 & 2.1 &    17.2 &    13.2 $\pm$ 1.8 &    6.8 &    2.8 & 1 & -- & --- & -- & -- & -- & FF UMa & star G5   \\ 
SAXWFC J0945.6-1419.7$^{j}$ & 09 45 35.0 & -14 19 44.0 & 1.9 &    66.9 &     7.0 $\pm$ 1.1 &    6.0 &    4.5 & 1 & 156.7 & --- & -- & -- & -- & NGC 2992 & AGN, Seyfert 1  \\ 
SAXWFC J0947.6-3056.7 & 09 47 35.3 & -30 56 43.8 & 1.9 &    76.8 &    10.4 $\pm$ 0.6 &   16.2 &    5.3 & 10 & 623.3 &  9.6 $\pm$  2.6 & 8.9 & 15.0 &  6.8 & MCG-5-23-16 & AGN, Seyfert 2  \\ 
SAXWFC J0955.5+6903.8 & 09 55 31.8 &  69 03 50.0 & 1.9 &    87.2 &     3.1 $\pm$ 0.4 &    7.2 &    3.6 & 4 & 356.1 &  3.8 $\pm$  1.5 & 4.0 &  6.4 &  3.1 & NGC 3031 & AGN, Seyfert 1  \\ 
SAXWFC J0955.6+6940.6 & 09 55 38.6 &  69 40 33.2 & 2.0 &    48.5 &     4.7 $\pm$ 0.4 &   10.4 &    2.8 & 3 & 217.6 &  4.8 $\pm$  0.7 & 1.7 &  5.8 &  4.4 & NGC 3034 & Galaxy   \\ 
SAXWFC J1009.7-5817.6 & 10 09 43.8 & -58 17 35.5 & 1.3 &    87.2 &    33.0 $\pm$ 1.0 &   41.7 &   13.3 & 4 & 335.0 & 13.1 $\pm$ 11.7 & 172.4 & 33.0 &  7.8 & GRO J1008-57 & Be HMXRB, XP, T  \\ 
SAXWFC J1037.6-5647.6 & 10 37 34.8 & -56 47 33.3 & 1.8 &    83.5 &     5.9 $\pm$ 0.5 &   11.6 &    6.8 & 3 & 368.8 &  6.4 $\pm$  0.9 & 1.5 &  7.7 &  5.9 & RX J1037.5-5647 & Be HMXRB, XP  \\ 
SAXWFC J1045.1-5941.1 & 10 45 03.2 & -59 41 08.5 & 1.4 &    54.5 &    15.7 $\pm$ 0.7 &   23.1 &    6.9 & 42 & 2879.5 &  9.3 $\pm$  4.1 & 11.1 & 21.0 &  5.4 & Eta Car & LBV  \\ 
SAXWFC J1052.6-6216.0 & 10 52 33.3 & -62 15 59.4 & 1.7 &    39.3 &    17.6 $\pm$ 1.6 &   10.6 &    9.9 & 1 & -- & --- & -- & -- & -- & 2E 1050.8-6200 & unclassified    \\ 
SAXWFC J1104.4+3812.0 & 11 04 23.0 &  38 12 00.3 & 1.3 &    71.4 &    53.0 $\pm$ 1.8 &   33.9 &    3.3 & 31 & 1517.1 & 21.2 $\pm$ 13.1 & 88.6 & 58.0 &  6.5 & MKN 421 & AGN, BL Lac   \\ 
SAXWFC J1106.8+7233.6 & 11 06 50.5 &  72 33 34.2 & 1.4 &    84.7 &     9.1 $\pm$ 0.6 &   14.0 &    3.6 & 19 & 1244.8 &  5.9 $\pm$  1.9 & 7.6 & 11.0 &  3.3 & NGC 3516 & AGN, Seyfert 1   \\ 
SAXWFC J1118.2+4802.9 & 11 18 14.9 &  48 02 55.6 & 1.3 &    79.2 &    69.2 $\pm$ 1.5 &   80.7 &   19.0 & 9 & 357.4 & 66.1 $\pm$  7.7 & 14.5 & 80.0 & 54.0 & XTE J1118+480 & LMXRB, BH    \\ 
SAXWFC J1121.3-6037.5 & 11 21 15.9 & -60 37 30.0 & 1.3 &    67.0 &   559.9 $\pm$ 11.0 &  545.7 &  209.8 & 152 & 8498.3 & 40.4 $\pm$ 170.5 & 736.5 & 610.0 &  8.0 & Cen X-3 & HMXRB, XP  \\ 
SAXWFC J1124.5-5915.7 & 11 24 31.6 & -59 15 44.6 & 1.8 &   147.8 &     4.2 $\pm$ 0.6 &    6.8 &    0.9 & 4 & 529.9 &  3.3 $\pm$  4.4 & 7.8 & 12.0 &  2.5 & SNR 292.0+01.8 & SNR    \\ 
SAXWFC J1138.9-3744.7 & 11 38 56.0 & -37 44 41.2 & 1.7 &   167.4 &     8.4 $\pm$ 0.8 &   10.7 &    3.5 & 12 & 975.7 &  7.1 $\pm$  2.0 & 4.6 & 11.0 &  4.2 & NGC 3783 & AGN, Seyfert 1  \\ 
SAXWFC J1139.5-6523.4 & 11 39 28.1 & -65 23 21.8 & 1.3 &    96.7 &    38.8 $\pm$ 1.0 &   60.1 &   11.9 & 10 & 876.2 & 14.0 $\pm$ 19.2 & 123.3 & 64.0 &  6.1 & HD 101379 & star G5III   \\ 
SAXWFC J1141.5-6411.3 & 11 41 29.6 & -64 11 18.2 & 2.3 &   130.0 &     6.0 $\pm$ 0.8 &    7.0 &    3.2 & 2 & 225.5 &  5.1 $\pm$  1.2 & 2.6 &  6.0 &  4.3 & V1033 Cen & CV, AM Her  \\ 
SAXWFC J1143.6+7141.1 & 11 43 35.2 &  71 41 04.5 & 1.7 &    60.5 &     4.5 $\pm$ 0.6 &    7.8 &    0.0 & 3 & 201.1 &  4.6 $\pm$  0.1 & 0.1 &  4.7 &  4.4 & YY Dra & CV    \\ 
SAXWFC J1143.8-6108.3 & 11 43 48.2 & -61 08 17.8 & 1.7 &    45.3 &    14.7 $\pm$ 1.0 &   14.2 &    9.2 & 1 & -- & --- & -- & -- & -- & IGR J11435-6109 & Be HMXRB, T  \\ 
SAXWFC J1147.4-6156.5 & 11 47 21.5 & -61 56 32.2 & 1.3 &    80.5 &    41.0 $\pm$ 1.3 &   35.7 &   23.5 & 71 & 5291.0 & 16.8 $\pm$  6.8 & 19.0 & 54.0 &  6.8 & 1E 1145.1-6141 & SG HMXRB, XP   \\ 
SAXWFC J1148.0-6211.9 & 11 48 02.4 & -62 11 51.3 & 1.3 &    17.9 &   108.8 $\pm$ 3.0 &   47.9 &   27.5 & 6 & 308.5 & 25.7 $\pm$ 38.4 & 365.6 & 110.0 & 11.0 & 4U 1145-619 & Be HMXRB, XP  \\

SAXWFC J1210.4+3924.4 & 12 10 25.2 &  39 24 26.2 & 1.3 &    66.5 &    12.7 $\pm$ 0.6 &   24.8 &   15.4 & 40 & 1666.5 & 14.8 $\pm$  6.5 & 22.1 & 33.0 &  8.0 & NGC 4151 & AGN, Seyfert 1  \\ 
SAXWFC J1213.2-6452.7 & 12 13 10.7 & -64 52 44.7 & 1.3 &    96.6 &    25.7 $\pm$ 0.9 &   36.3 &   13.6 & 20 & 1099.7 & 18.5 $\pm$  7.6 & 34.3 & 34.0 &  8.3 & 1ES 1210-64.6 & unclassified    \\ 
SAXWFC J1226.6-6245.7 & 12 26 36.1 & -62 45 43.2 & 1.3 &    22.8 &   173.4 $\pm$ 3.6 &  175.5 &   76.6 & 105 & 6548.9 & 37.3 $\pm$ 53.7 & 314.2 & 340.0 &  8.1 & GX 301-2 & SG HMXRB, XP   \\ 
SAXWFC J1229.1+0202.3 & 12 29 05.4 &  02 02 20.7 & 1.9 &    22.5 &     7.9 $\pm$ 0.9 &    8.1 &    0.6 & 3 & 32.3 &  9.8 $\pm$  6.7 & 10.7 & 21.0 &  7.9 & 3C 273 & AGN, BL Lac   \\ 
SAXWFC J1248.9-4119.6 & 12 48 54.7 & -41 19 35.4 & 1.6 &    97.6 &     5.8 $\pm$ 0.7 &    8.2 &    1.2 & 1 & -- & --- & -- & -- & -- & ABELL 3526 & ClG   \\ 
SAXWFC J1249.7-5905.1 & 12 49 42.0 & -59 05 04.2 & 1.3 &    52.2 &    47.8 $\pm$ 1.2 &   61.7 &   13.4 & 50 & 3436.9 & 22.1 $\pm$ 15.5 & 72.2 & 73.0 &  9.1 & 4U 1246-58 & LMXRB, B    \\ 
SAXWFC J1252.4-2914.0 & 12 52 21.6 & -29 13 58.8 & 1.5 &    32.7 &    10.7 $\pm$ 0.7 &   15.5 &    4.4 & 8 & 545.9 & 10.4 $\pm$  1.4 & 1.1 & 14.0 &  9.4 & EX Hya & CV    \\ 
SAXWFC J1257.6-6917.0 & 12 57 35.3 & -69 17 00.9 & 1.3 &   135.7 &    78.1 $\pm$ 1.6 &  116.7 &   12.5 & 178 & 12118.0 & 69.5 $\pm$ 12.4 & 18.3 & 130.0 & 38.0 & 4U 1254-690 & LMXRB, B, D  \\ 
SAXWFC J1259.5+2756.6 & 12 59 31.7 &  27 56 34.4 & 1.5 &    75.4 &     9.8 $\pm$ 1.1 &    8.5 &    1.2 & 1 & -- & --- & -- & -- & -- & Coma Cluster & ClG  \\ 
SAXWFC J1325.5-4302.1 & 13 25 29.4 & -43 02 03.1 & 1.4 &   148.8 &    20.5 $\pm$ 0.7 &   34.9 &   18.5 & 32 & 1863.7 & 23.2 $\pm$  5.3 & 16.1 & 33.0 & 13.0 & Cen A & AGN, Seyfert 2  \\ 
SAXWFC J1326.7-6208.8 & 13 26 42.7 & -62 08 47.0 & 1.4 &   124.3 &    36.8 $\pm$ 1.6 &   22.0 &    1.4 & 33 & 2724.2 & 14.7 $\pm$ 12.9 & 31.9 & 59.0 &  8.0 & 4U 1323-619 & LMXRB, B, D  \\ 
SAXWFC J1336.0-3417.1 & 13 35 57.2 & -34 17 07.0 & 1.6 &   141.9 &     6.2 $\pm$ 0.5 &   11.0 &    3.9 & 1 & -- & --- & -- & -- & -- & MCG-06-30-015 & AGN, Seyfert 1  \\ 
SAXWFC J1347.4-3252.0 & 13 47 25.4 & -32 51 59.0 & 1.6 &   140.8 &     7.0 $\pm$ 0.7 &    9.4 &    2.6 & 4 & 356.9 &  7.8 $\pm$  0.8 & 0.9 &  9.0 &  7.0 & ABELL 3571 & ClG  \\ 
SAXWFC J1348.8+2635.9 & 13 48 50.9 &  26 35 53.5 & 1.6 &    78.2 &     6.3 $\pm$ 0.7 &    8.2 &    3.1 & 4 & 148.9 &  6.8 $\pm$  0.8 & 0.5 &  8.1 &  6.3 & PKS 1346+26 & AGN, RG  \\ 
SAXWFC J1349.3-3019.1 & 13 49 17.6 & -30 19 07.3 & 1.4 &    96.4 &    18.4 $\pm$ 0.9 &   18.9 &    5.9 & 14 & 780.2 & 14.5 $\pm$  2.5 & 4.0 & 19.0 & 11.0 & ESO 445-50 & AGN, Seyfert 1  \\

SAXWFC J1358.0-6443.7 & 13 58 03.0 & -64 43 39.0 & 1.3 &    24.8 &    84.3 $\pm$ 2.5 &   41.2 &   15.8 & 7 & 269.6 & 55.9 $\pm$ 23.0 & 84.4 & 84.0 & 25.0 & Cen X-2 & LMXRB, BH, T  \\ 
SAXWFC J1413.3-0311.7 & 14 13 15.4 & -03 11 44.5 & 1.5 &    58.5 &    10.4 $\pm$ 0.6 &   19.0 &    6.0 & 8 & 429.8 &  9.2 $\pm$  2.2 & 3.8 & 13.0 &  6.3 & MKN 1376 & AGN, Seyfert 2  \\ 
SAXWFC J1421.2-6241.7 & 14 21 12.3 & -62 41 42.0 & 1.5 &    33.2 &    17.0 $\pm$ 1.2 &   15.2 &   15.8 & 13 & 808.8 & 18.7 $\pm$  9.2 & 17.0 & 37.0 &  9.5 & 4U 1416-62 & Be HMXRB, XP, T  \\ 
SAXWFC J1428.6-5421.9 & 14 28 37.2 & -54 21 56.8 & 1.5 &   107.5 &    32.3 $\pm$ 1.9 &   16.1 &    6.7 & 2 & 210.5 & 27.2 $\pm$  6.5 & 13.4 & 32.0 & 23.0 & SAX J1428.6-5422 & T, unclassified    \\ 

SAXWFC J1513.9-5908.8 & 15 13 55.3 & -59 08 49.5 & 1.7 &    56.3 &     9.5 $\pm$ 1.5 &    6.2 &    6.0 & 1 & -- & --- & -- & -- & -- & PSR B1509-58 & RP   \\ 
SAXWFC J1520.6-5709.8 & 15 20 38.3 & -57 09 50.4 & 1.3 &    58.0 &  2020.2 $\pm$ 40.1 &  537.5 &  190.9 & 95 & 5318.1 & 1563.6 $\pm$ 1072.0 & 631.4 & 5700.0 & 530.0 & Cir X-1 & XB, B, T  \\ 
SAXWFC J1542.4-5222.7 & 15 42 26.1 & -52 22 43.6 & 1.4 &    58.7 &    58.2 $\pm$ 3.0 &   20.1 &   28.7 & 20 & 1508.8 & 35.6 $\pm$ 14.6 & 13.8 & 72.0 & 22.0 & 4U 1538-522 & SG HMXRB, XP   \\ 
\hline

\end{tabular}
\end{minipage}
\end{table*}

\newpage
\thispagestyle{empty}
\markright{}
\authorrunning{}
\titlerunning{}

\begin{table*}[tbh]
\setcounter{table}{2}

\scriptsize

\begin{minipage}{250mm}
\caption{ BeppoSAX WFC X-ray Source Catalogue}

\begin{tabular}{rccccrrrrrrlllll}
\hline
\hline
 WFC Name & RA(2000) & DEC(2000) & $^{a}$ERR$_{RAD}$ & EXPO &
   $^{b}$Flux$_{2-10}$ & $^{c}$SNR1 & $^{d}$SNR2 & N$_{det}$ &
   EXPO$_{total}$ & $^{e}<$FLUX$>_{2-10}$ & $\chi^{2}_{flux}$  &
   $^{f}$Flux$_{MAX}$  & $^{f}$Flux$_{MIN }$ & Name & Class   \\
   & hh mm ss & o  '  "   & arcmin & ks~~~ & &  &  &   & ks ~~~~~ &  & &  & &  &
  \\
\hline
SAXWFC J1544.0-5646.5 & 15 44 01.8 & -56 46 32.5 & 1.4 &   152.5 &    16.1 $\pm$ 2.0 &    7.8 &   12.0 & 1 & -- & --- & -- & -- & -- & XTE J1543-568 & Be HMXRB, XP, T    \\ 
SAXWFC J1547.9-6234.6 & 15 47 55.5 & -62 34 34.6 & 1.3 &   121.3 &    90.0 $\pm$ 2.3 &   64.6 &   32.1 & 86 & 5519.3 & 70.2 $\pm$ 20.4 & 19.1 & 150.0 & 33.0 & 4U 1543-624 & LMXRB    \\ 
SAXWFC J1550.9-5628.5 & 15 50 52.8 & -56 28 27.8 & 1.3 &    35.4 &  5078.8 $\pm$ 99.3 &  305.0 &   40.8 & 20 & 1394.6 & 83.6 $\pm$ 2523.5 & 1770.5 & 6100.0 & 41.0 & XTE J1550-564 & LMXRB, BH, T    \\ 
SAXWFC J1558.3+2712.7 & 15 58 19.4 &  27 12 43.9 & 1.6 &    80.0 &     6.5 $\pm$ 0.7 &    9.1 &    3.4 & 8 & 394.6 &  7.0 $\pm$  2.5 & 6.8 & 11.0 &  4.1 & ABELL 2142 & ClG  \\ 
SAXWFC J1601.1-6044.3 & 16 01 06.1 & -60 44 20.4 & 1.4 &    60.4 &    40.7 $\pm$ 1.5 &   32.0 &    6.2 & 47 & 3400.5 & 41.3 $\pm$  6.8 & 2.6 & 61.0 & 31.0 & 4U 1556-60 & LMXRB    \\ 
SAXWFC J1612.7-5224.6 & 16 12 39.5 & -52 24 36.0 & 1.3 &    40.8 &  1146.4 $\pm$ 23.2 &  222.5 &  115.4 & 18 & 704.3 & 104.3 $\pm$ 322.3 & 280.8 & 1100.0 & 44.0 & 4U 1608-522 & LMXRB, B, T \\ 

SAXWFC J1619.9-1537.4 & 16 19 54.1 & -15 37 25.3 & 1.3 &     6.9 & 21072.2 $\pm$ 414.7 &  570.6 &  111.1 & 99 & 1974.0 & 24754.9 $\pm$ 4116.0 & 48.6 & 40000.0 & 19000.0 & Sco X-1 & LMXRB   \\ 
SAXWFC J1628.0-4912.0 & 16 28 00.6 & -49 12 02.8 & 1.3 &   131.1 &   125.3 $\pm$ 3.4 &   50.4 &   27.3 & 54 & 2920.1 & 115.9 $\pm$ 17.1 & 5.0 & 170.0 & 77.0 & 4U 1624-490 & LMXRB, D   \\ 
SAXWFC J1628.6+3931.5 & 16 28 37.8 &  39 31 27.8 & 1.6 &    78.9 &     7.0 $\pm$ 0.8 &    8.6 &    2.8 & 16 & 761.1 &  6.0 $\pm$  1.7 & 2.7 & 10.0 &  4.6 & ABELL 2199 & ClG  \\ 
SAXWFC J1632.1-6727.7 & 16 32 07.8 & -67 27 40.3 & 1.5 &    45.3 &    16.6 $\pm$ 0.8 &   20.0 &   11.8 & 66 & 4581.5 & 16.5 $\pm$  3.2 & 3.4 & 27.0 &  9.8 & 4U 1627-673 & LMXRB, XP   \\ 
SAXWFC J1634.0-4722.8 & 16 33 59.0 & -47 22 50.8 & 1.3 &    39.4 &   564.4 $\pm$ 12.1 &   98.7 &   55.6 & 9 & 646.2 & 124.4 $\pm$ 210.8 & 415.0 & 590.0 & 69.0 & 4U 1630-472 & LMXRB, BH  \\ 
SAXWFC J1638.1-6422.2 & 16 38 04.9 & -64 22 13.8 & 1.7 &    68.9 &     9.8 $\pm$ 1.2 &    7.2 &    0.0 & 2 & 73.8 & 11.2 $\pm$  7.3 & 9.8 & 20.0 &  9.8 & IGR J16377-6423 & ClG   \\ 
SAXWFC J1641.0-5345.2 & 16 40 57.7 & -53 45 11.5 & 1.3 &    51.7 &   634.3 $\pm$ 12.8 &  214.2 &   65.2 & 69 & 3285.2 & 32.9 $\pm$ 107.5 & 1332.5 & 700.0 &  6.4 & 4U 1636-536 & LMXRB, B   \\ 

SAXWFC J1645.8-4536.3 & 16 45 45.5 & -45 36 18.7 & 1.3 &    56.1 &   960.3 $\pm$ 19.1 &  280.0 &   97.1 & 129 & 6769.2 & 25.0 $\pm$ 151.9 & 1588.0 & 1300.0 & 10.0 & GX 340+0 & LMXRB    \\ 
SAXWFC J1653.8+3944.2 & 16 53 47.4 &  39 44 15.0 & 1.5 &    23.5 &    21.9 $\pm$ 0.9 &   28.1 &   11.0 & 70 & 3245.3 & 10.3 $\pm$  8.2 & 43.8 & 42.0 &  4.2 & MKN 501 & AGN, BL Lac   \\ 
SAXWFC J1654.0-3950.9 & 16 53 57.9 & -39 50 52.0 & 1.3 &    14.6 &  6810.3 $\pm$ 133.3 &  504.5 &  167.6 & 42 & 1448.0 & 59.9 $\pm$ 1876.6 & 2401.7 & 7400.0 & 43.0 & GRO J1655-40 & LMXRB, BH, T  \\ 
SAXWFC J1657.9+3521.1 & 16 57 51.6 &  35 21 03.2 & 1.3 &    72.3 &    98.7 $\pm$ 2.0 &  177.0 &   68.7 & 42 & 2061.1 & 18.4 $\pm$ 70.5 & 524.7 & 240.0 &  4.1 & Her X-1 & LMXRB, XP \\ 
SAXWFC J1700.7-4139.4 & 17 00 43.0 & -41 39 21.2 & 1.6 &    72.8 &   112.7 $\pm$ 8.5 &   12.0 &   11.2 & 13 & 695.4 & 61.3 $\pm$ 28.0 & 14.7 & 120.0 & 30.0 & OAO 1657-415 & SG HMXRB, XP   \\ 
SAXWFC J1702.1-2957.9 & 17 02 05.9 & -29 57 56.1 & 1.8 &    74.6 &    76.8 $\pm$ 4.7 &   15.9 &    6.6 & 25 & 1452.3 & 64.9 $\pm$  7.5 & 2.0 & 77.0 & 49.0 & XB 1658-298 & LMXRB, B, T, E  \\ 
SAXWFC J1702.9-4847.5 & 17 02 51.1 & -48 47 30.4 & 1.3 &    84.6 &   481.5 $\pm$ 10.1 &  100.9 &    3.7 & 19 & 736.1 & 123.2 $\pm$ 205.1 & 362.9 & 600.0 & 41.0 & GX 339-4 & LMXRB, BH  \\ 

SAXWFC J1703.9-3750.9 & 17 03 54.5 & -37 50 56.7 & 1.3 &    86.9 &   286.8 $\pm$ 7.9 &   48.8 &   76.7 & 96 & 4763.3 & 137.8 $\pm$ 81.5 & 80.2 & 440.0 & 46.0 & 4U 1700-377 & SG HMXRB   \\ 
SAXWFC J1704.2+7838.0 & 17 04 11.3 &  78 38 02.0 & 1.5 &   206.7 &     4.2 $\pm$ 0.5 &    8.6 &    2.4 & 11 & 977.4 &  4.4 $\pm$  6.8 & 6.2 & 21.0 &  3.0 & NGC 6331 & Galaxy    \\ 
SAXWFC J1705.7-3625.1 & 17 05 44.4 & -36 25 03.3 & 1.3 &    17.5 &  1393.4 $\pm$ 27.9 &  355.2 &  103.6 & 154 & 6154.0 & 36.2 $\pm$ 247.4 & 2058.0 & 2100.0 & 14.0 & GX 349+2 & LMXRB    \\ 
SAXWFC J1706.2-4302.0 & 17 06 11.9 & -43 01 58.0 & 1.3 &    37.2 &   202.8 $\pm$ 5.8 &   46.7 &   37.2 & 81 & 4513.3 & 89.7 $\pm$ 45.0 & 36.8 & 260.0 & 43.0 & 4U 1702-429 & LMXRB, B    \\ 
SAXWFC J1706.6+2357.5 & 17 06 35.9 &  23 57 27.0 & 1.7 &     5.1 &    26.8 $\pm$ 2.5 &   10.2 &    2.7 & 5 & 87.7 & 16.0 $\pm$  6.4 & 12.8 & 27.0 & 11.0 & 4U 1700+24 & LMXRB    \\ 
SAXWFC J1708.9-4406.9 & 17 08 52.1 & -44 06 53.2 & 1.3 &    18.3 &   698.9 $\pm$ 14.2 &  160.3 &   54.4 & 114 & 5401.1 &  8.5 $\pm$ 227.8 & 968.3 & 920.0 &  2.0 & 4U 1705-440 & LMXRB, B   \\ 
SAXWFC J1711.6-3807.3 & 17 11 34.2 & -38 07 17.7 & 1.3 &    86.2 &   108.2 $\pm$ 4.4 &   26.7 &   27.6 & 5 & 320.5 & 99.2 $\pm$ 12.4 & 4.7 & 110.0 & 79.0 & SAX J1711.6-3808 & LMXRB, BH, T  \\ 
SAXWFC J1712.4-4050.7 & 17 12 21.3 & -40 50 43.0 & 1.4 &    86.6 &    94.3 $\pm$ 3.8 &   27.6 &    9.8 & 33 & 1744.4 & 71.8 $\pm$ 13.7 & 5.1 & 100.0 & 51.0 & 4U 1708-40 & LMXRB, B   \\ 
SAXWFC J1712.6-3739.5 & 17 12 35.4 & -37 39 31.3 & 1.4 &    34.5 &    76.6 $\pm$ 5.2 &   13.8 &    2.2 & 8 & 354.2 & 56.5 $\pm$ 11.1 & 5.8 & 77.0 & 43.0 & SAX J1712.6-3739 & LMXRB, B, T \\ 
SAXWFC J1714.3-3402.8 & 17 14 18.5 & -34 02 48.8 & 1.4 &    40.5 &    87.7 $\pm$ 4.8 &   18.6 &    7.7 & 12 & 667.9 & 79.3 $\pm$ 14.8 & 5.1 & 100.0 & 62.0 & 4U 1711-34 & LMXRB, B, T    \\ 
SAXWFC J1716.0-3851.9 & 17 15 59.8 & -38 51 53.6 & 1.4 &    17.7 &    62.7 $\pm$ 4.2 &   15.4 &    3.9 & 43 & 1993.8 & 62.8 $\pm$ 13.6 & 3.6 & 100.0 & 40.0 & XTE J1716-389 & SG HMXRB   \\ 

SAXWFC J1723.6-3739.4 & 17 23 37.2 & -37 39 24.1 & 1.3 &    36.0 &   125.1 $\pm$ 4.4 &   32.9 &   17.0 & 9 & 511.8 & 123.3 $\pm$ 28.6 & 23.6 & 150.0 & 74.0 & XTE J1723-376 & LMXRB, B, T   \\ 
SAXWFC J1727.5-3048.1 & 17 27 29.4 & -30 48 04.6 & 1.4 &    50.4 &    76.5 $\pm$ 4.7 &   16.5 &   12.3 & 74 & 3990.1 & 51.4 $\pm$ 20.2 & 6.4 & 170.0 & 34.0 & 4U 1724-307 & LMXRB, GlC, B  \\ 
SAXWFC J1728.6+5901.5 & 17 28 34.0 &  59 01 29.2 & 1.4 &    15.1 &    38.9 $\pm$ 1.5 &   27.7 &    5.1 & 5 & 166.0 & 15.0 $\pm$ 14.2 & 157.3 & 39.0 &  6.8 & GR Dra & Star G0   \\ 
SAXWFC J1731.7-1657.6 & 17 31 42.8 & -16 57 38.8 & 1.3 &    35.6 &   473.3 $\pm$ 10.4 &   81.9 &   26.5 & 163 & 5076.0 & 501.8 $\pm$ 76.5 & 11.1 & 980.0 & 350.0 & GX 9+9 & LMXRB    \\ 
SAXWFC J1731.9-3350.2 & 17 31 55.1 & -33 50 10.3 & 1.3 &    60.4 &   356.4 $\pm$ 8.2 &   79.5 &   64.9 & 135 & 5331.8 & 10.7 $\pm$ 75.3 & 534.5 & 440.0 &  1.5 & GX 354-0 & LMXRB, B   \\ 
SAXWFC J1732.0-2444.0 & 17 32 01.1 & -24 44 02.0 & 1.4 &    41.1 &    89.5 $\pm$ 5.4 &   16.8 &   41.4 & 24 & 1180.4 & 51.4 $\pm$ 17.9 & 9.8 & 89.0 & 33.0 & GX 1+4 & LMXRB, XP   \\ 
SAXWFC J1732.5+7413.2 & 17 32 31.0 &  74 13 14.1 & 2.4 &    33.5 &     4.4 $\pm$ 0.5 &    7.8 &    0.0 & 1 & -- & --- & -- & -- & -- & DR Dra & RSCVn  \\ 
SAXWFC J1733.4-3323.8 & 17 33 22.7 & -33 23 47.7 & 1.3 &    55.5 &   479.3 $\pm$ 10.5 &   98.6 &   78.5 & 24 & 924.0 & 192.6 $\pm$ 120.8 & 211.8 & 480.0 & 47.0 & Rapid Burster & LMXRB, GlC, B, T  \\ 
SAXWFC J1734.2-2605.2 & 17 34 10.5 & -26 05 13.5 & 1.3 &   126.0 &   373.2 $\pm$ 8.1 &  105.3 &   39.2 & 115 & 4500.5 & 23.6 $\pm$ 149.1 & 625.9 & 630.0 &  5.3 & KS 1731-260 & LMXRB, B, T   \\ 
SAXWFC J1738.2-2659.5 & 17 38 13.2 & -26 59 27.6 & 1.5 &    15.0 &    58.0 $\pm$ 5.8 &   10.1 &    2.2 & 9 & 465.7 & 24.6 $\pm$ 12.4 & 6.8 & 58.0 & 18.0 & SLX 1735-269 & LMXRB, B   \\ 
SAXWFC J1738.9-4426.8 & 17 38 56.7 & -44 26 48.4 & 1.3 &   129.2 &   410.0 $\pm$ 8.2 &  167.9 &   59.2 & 115 & 5856.2 & 16.5 $\pm$ 85.4 & 1206.4 & 520.0 &  3.5 & 4U 1735-444 & LMXRB, B   \\ 
SAXWFC J1740.1-3103.0$^{i}$ & 17 40 07.6 & -31 03 00.0 & 1.6 &   127.5 &    28.5 $\pm$ 4.8 &    5.8 &    8.5 & 6 & 704.9 & 24.1 $\pm$ 24.8 & 5.3 & 81.0 & 11.0 & GRS 1737-310 & LMXRB, T   \\ 
SAXWFC J1740.7-2818.5 & 17 40 40.8 & -28 18 30.2 & 1.6 &    17.5 &   123.7 $\pm$ 9.3 &   12.4 &   10.1 & 1 & -- & --- & -- & -- & -- & SLX 1737-282 & LMXRB, B \\ 
SAXWFC J1742.7-2744.9 & 17 42 39.6 & -27 44 51.7 & 1.3 &   151.1 &   360.7 $\pm$ 8.6 &   74.6 &    2.7 & 26 & 636.6 & 231.5 $\pm$ 79.2 & 96.5 & 370.0 & 80.0 & GRS 1739-278 & LMXRB, BH, T \\ 
SAXWFC J1743.8-2945.1 & 17 43 48.5 & -29 45 03.9 & 1.7 &    60.8 &    35.5 $\pm$ 3.2 &   11.4 &   20.9 & 50 & 2914.5 & 30.2 $\pm$  6.9 & 2.0 & 63.0 & 20.0 & 1E 1740.7-2942 & LMXRB, BH \\ 
SAXWFC J1744.5-2844.6 & 17 44 31.6 & -28 44 35.5 & 1.3 &   126.8 &   353.2 $\pm$ 8.4 &   76.2 &  154.8 & 20 & 1181.2 & 100.7 $\pm$ 174.7 & 221.1 & 670.0 & 32.0 & GRO J1744-28 & LMXRB, XP, B, T \\ 
SAXWFC J1744.9-2921.2 & 17 44 52.0 & -29 21 10.8 & 2.0 &    55.2 &    26.5 $\pm$ 4.2 &    6.4 &    8.6 & 3 & 174.7 & 29.7 $\pm$  4.4 & 1.5 & 34.0 & 27.0 & KS 1741-293 & LMXRB, B, T    \\ 
SAXWFC J1746.0-2931.0 & 17 46 01.6 & -29 30 59.0 & 1.4 &    56.5 &    69.9 $\pm$ 3.1 &   24.6 &   20.4 & 57 & 3236.2 & 39.8 $\pm$ 16.1 & 16.2 & 84.0 & 23.0 & 1A 1742-294 & LMXRB, B  \\ 
SAXWFC J1747.0-2853.3 & 17 46 59.2 & -28 53 17.8 & 1.3 &    79.2 &   242.2 $\pm$ 6.2 &   61.7 &   31.4 & 16 & 754.5 & 75.3 $\pm$ 69.7 & 141.7 & 240.0 & 27.0 & SAX J1747.0-2853 & LMXRB, B, T   \\ 
SAXWFC J1747.5-3001.5 & 17 47 28.1 & -30 01 28.2 & 1.3 &    60.2 &    71.6 $\pm$ 4.0 &   19.2 &   14.0 & 58 & 3353.4 & 46.3 $\pm$ 12.3 & 8.1 & 77.0 & 26.0 & SLX 1744-300 & LMXRB, B    \\ 
SAXWFC J1747.9-3009.0 & 17 47 54.1 & -30 08 58.5 & 1.8 &     6.8 &   107.6 $\pm$ 14.9 &    6.6 &    1.0 & 0 & -- & --- & -- & -- & -- & 1RXS J174735.7-300953 & unclassified   \\ 
SAXWFC J1747.9-2633.3 & 17 47 56.8 & -26 33 16.5 & 1.3 &    63.4 &   867.3 $\pm$ 17.5 &  241.2 &  114.6 & 148 & 5428.6 & 31.6 $\pm$ 175.7 & 1634.6 & 1500.0 &  6.7 & GX 3+1 & LMXRB, B   \\ 
SAXWFC J1748.1-2446.1 & 17 48 06.4 & -24 46 08.7 & 1.3 &    41.2 &   866.2 $\pm$ 18.0 &  149.8 &  115.7 & 13 & 587.2 & 240.7 $\pm$ 262.2 & 644.9 & 890.0 & 37.0 & EXO 1745-248 & LMXRB, GlC, B, T  \\ 
SAXWFC J1748.9-2021.3 & 17 48 53.8 & -20 21 17.6 & 1.3 &    20.9 &   336.5 $\pm$ 8.5 &   62.1 &   25.1 & 5 & 156.3 & 133.2 $\pm$ 144.3 & 386.5 & 340.0 & 38.0 & 4U 1745-203 & LMXRB, GlC, T, B  \\ 
\hline

\end{tabular}
\end{minipage}
\end{table*}

\newpage
\thispagestyle{empty}
\markright{}

\begin{table*}[tbh]
\setcounter{table}{2}

\scriptsize

\begin{minipage}{250mm}
\caption{ BeppoSAX WFC X-ray Source Catalogue}

\begin{tabular}{rccccrrrrrrlllll}
\hline
\hline
 WFC Name & RA(2000) & DEC(2000) & $^{a}$ERR$_{RAD}$ & EXPO &
   $^{b}$Flux$_{2-10}$ & $^{c}$SNR1 & $^{d}$SNR2 & N$_{det}$ &
   EXPO$_{total}$ & $^{e}<$FLUX$>_{2-10}$ & $\chi^{2}_{flux}$  &
   $^{f}$Flux$_{MAX}$  & $^{f}$Flux$_{MIN }$ & Name & Class   \\
   & hh mm ss & o  '  "   & arcmin & ks~~~ & &  &  &   & ks ~~~~~ &  & &  & &  &
  \\
\hline
SAXWFC J1750.2-3703.1 & 17 50 09.4 & -37 03 07.5 & 1.4 &    40.7 &    99.8 $\pm$ 4.4 &   24.6 &   15.3 & 71 & 3785.7 & 82.6 $\pm$ 17.8 & 9.0 & 130.0 & 46.0 & 4U 1746-370 & LMXRB, GlC, B, D  \\ 
SAXWFC J1750.4-2902.6 & 17 50 23.5 & -29 02 35.1 & 1.4 &   127.0 &    74.3 $\pm$ 4.0 &   19.6 &    8.9 & 3 & 198.8 & 81.3 $\pm$ 45.9 & 35.8 & 150.0 & 72.0 & SAX J1750.8-2900 & LMXRB, B, T   \\ 
SAXWFC J1750.8-3116.5 & 17 50 47.5 & -31 16 28.2 & 1.4 &    42.3 &    58.0 $\pm$ 3.5 &   17.1 &   12.5 & 18 & 977.4 & 47.6 $\pm$ 14.1 & 11.5 & 66.0 & 24.0 & GRS 1747-312 & LMXRB, GlC, B, T  \\ 
SAXWFC J1751.2-3037.7 & 17 51 10.0 & -30 37 41.1 & 1.4 &    15.9 &    89.6 $\pm$ 4.6 &   20.7 &   19.1 & 1 & -- & --- & -- & -- & -- & XTE J1751-305 & XB, XP   \\ 
SAXWFC J1755.5-3229.7 & 17 55 32.8 & -32 29 42.7 & 1.5 &    22.9 &    66.8 $\pm$ 4.4 &   15.1 &    0.5 & 2 & 90.9 & 58.1 $\pm$ 10.2 & 6.6 & 67.0 & 52.0 & XTE J1755-324 & LMXRB, BH, T    \\ 
SAXWFC J1801.1-2504.9 & 18 01 05.5 & -25 04 56.6 & 1.3 &    64.6 &  2044.4 $\pm$ 40.4 &  411.3 &  105.4 & 147 & 5325.7 & 46.7 $\pm$ 378.3 & 2350.5 & 4700.0 & 18.0 & GX 5-1 & LMXRB   \\ 
SAXWFC J1801.2-2544.4 & 18 01 13.1 & -25 44 25.0 & 1.7 &    67.3 &    49.7 $\pm$ 4.1 &   12.1 &   11.2 & 20 & 1322.8 & 45.1 $\pm$  9.2 & 2.2 & 66.0 & 34.0 & GRS 1758-258 & LMXRB, BH  \\ 
SAXWFC J1801.5-2031.6 & 18 01 31.9 & -20 31 34.3 & 1.3 &    31.0 &  1151.8 $\pm$ 23.6 &  220.5 &  122.7 & 143 & 5084.5 & 135.1 $\pm$ 199.0 & 1921.3 & 2500.0 &  8.9 & GX 9+1 & LMXRB    \\ 
SAXWFC J1805.5-2031.1 & 18 05 30.5 & -20 31 06.2 & 1.6 &    20.6 &    84.6 $\pm$ 6.7 &   12.1 &    1.1 & 3 & 158.6 & 66.3 $\pm$ 16.2 & 5.8 & 85.0 & 56.0 & SAX J1805.5-2031 & T    \\ 
SAXWFC J1806.9-2434.7 & 18 06 51.3 & -24 34 44.0 & 2.3 &    58.1 &    31.7 $\pm$ 5.0 &    6.1 &    6.8 & 1 & -- & --- & -- & -- & -- & 2S 1803-245 & LMXRB, B, T    \\ 
SAXWFC J1808.4-3659.0 & 18 08 26.1 & -36 59 01.3 & 1.3 &     7.9 &   177.9 $\pm$ 5.2 &   43.7 &   14.5 & 14 & 118.3 & 159.7 $\pm$ 56.4 & 31.8 & 230.0 & 57.0 & SAX J1808.4-3658 & LMXRB, XP, B, T \\ 
SAXWFC J1810.8-2609.0 & 18 10 48.8 & -26 09 02.5 & 2.0 &    57.7 &    40.0 $\pm$ 6.3 &    6.1 &    4.1 & 1 & -- & --- & -- & -- & -- & SAX J1810.8-2609 & LMXRB, B, T \\ 
SAXWFC J1814.5-1708.8 & 18 14 28.1 & -17 08 48.8 & 1.3 &    58.3 &   831.6 $\pm$ 17.3 &  146.9 &   41.6 & 161 & 4931.2 & 737.2 $\pm$ 96.9 & 162.2 & 16.2 & 530.0 & GX 13+1 & LMXRB, B   \\ 
SAXWFC J1815.1-1205.6 & 18 15 08.2 & -12 05 38.0 & 1.7 &    35.5 &    40.2 $\pm$ 4.2 &    9.1 &    3.8 & 9 & 252.1 & 37.4 $\pm$  5.6 & 1.3 & 48.0 & 30.0 & 4U 1812-12 & LMXRB, B   \\ 
SAXWFC J1816.0-1402.6 & 18 16 01.0 & -14 02 34.8 & 1.3 &    32.6 &  3580.1 $\pm$ 70.7 &  300.0 &   85.3 & 175 & 5293.6 & 1371.1 $\pm$ 217.3 & 24.0 & 3600.0 & 1100.0 & GX 17+2 & LMXRB, B   \\ 
SAXWFC J1816.4+4953.1 & 18 16 21.0 &  49 53 04.2 & 1.4 &   224.8 &    14.3 $\pm$ 0.6 &   22.5 &    5.6 & 71 & 4007.7 & 10.3 $\pm$  3.5 & 14.0 & 20.0 &  3.6 & AM Her & CV, AM Her  \\ 
SAXWFC J1819.4-2524.0 & 18 19 21.9 & -25 24 00.3 & 1.5 &    69.3 &    72.9 $\pm$ 8.9 &    7.8 &    8.1 & 1 & -- & --- & -- & -- & -- & SAX J1819.3-2525 & LMXRB, BH, T \\ 
SAXWFC J1821.9+6419.6 & 18 21 53.4 &  64 19 36.4 & 1.7 &    35.7 &     4.5 $\pm$ 0.6 &    7.0 &    1.7 & 3 & 202.9 &  3.1 $\pm$  1.1 & 5.4 &  4.5 &  2.4 & QSO B1821+643 & AGN, Seyfert 1  \\ 
SAXWFC J1823.7-3021.6 & 18 23 42.1 & -30 21 38.8 & 1.3 &   312.3 &   818.0 $\pm$ 16.5 &  168.2 &   75.3 & 134 & 5360.6 & 27.6 $\pm$ 207.6 & 1338.6 & 940.0 &  4.3 & 4U 1820-303 & LMXRB, GlC, B  \\ 
SAXWFC J1825.2-0001.0 & 18 25 14.2 & -00 00 58.6 & 1.4 &    15.5 &    74.5 $\pm$ 2.3 &   41.1 &    9.9 & 69 & 1318.6 & 67.1 $\pm$ 12.8 & 6.8 & 120.0 & 45.0 & 4U 1822-000 & LMXRB    \\ 
SAXWFC J1825.8-3707.0 & 18 25 46.0 & -37 06 58.3 & 1.5 &    12.9 &    51.3 $\pm$ 2.5 &   20.2 &   11.2 & 61 & 3140.8 & 50.0 $\pm$  7.8 & 1.9 & 72.0 & 39.0 & 4U 1822-371 & LMXRB, E  \\ 
SAXWFC J1829.6-2347.7 & 18 29 34.5 & -23 47 39.4 & 1.4 &    36.5 &    82.7 $\pm$ 3.8 &   21.0 &    8.7 & 58 & 2554.4 & 73.7 $\pm$ 15.6 & 3.3 & 120.0 & 54.0 & GS 1826-238 & LMXRB, B   \\ 
SAXWFC J1833.9+5143.5 & 18 33 53.2 &  51 43 28.2 & 1.5 &    25.8 &    11.4 $\pm$ 0.9 &   13.1 &    2.3 & 1 & -- & --- & -- & -- & -- & BY Dra & BY Dra   \\ 
SAXWFC J1835.0+3241.6 & 18 35 01.8 &  32 41 37.6 & 1.5 &   128.1 &     5.5 $\pm$ 0.5 &   11.4 &    2.0 & 4 & 392.8 &  6.0 $\pm$  0.5 & 0.8 &  6.7 &  5.5 & 3C 382 & AGN, Seyfert 1   \\ 
SAXWFC J1835.6-3258.2 & 18 35 38.1 & -32 58 14.5 & 1.7 &    35.2 &    27.2 $\pm$ 3.5 &    7.5 &    4.7 & 3 & 76.0 & 21.9 $\pm$  4.0 & 1.7 & 27.0 & 19.0 & XB 1832-330 & LMXRB, GlC, B  \\ 
SAXWFC J1839.9+0501.9 & 18 39 54.7 &  05 01 51.2 & 1.3 &    46.5 &   530.8 $\pm$ 10.7 &  372.5 &   88.4 & 94 & 1830.7 & 480.0 $\pm$ 100.3 & 36.8 & 1100.0 & 240.0 & Ser X-1 & LMXRB, B   \\ 
SAXWFC J1842.2+7945.6 & 18 42 13.2 &  79 45 38.1 & 1.6 &    62.1 &     5.6 $\pm$ 0.6 &    9.7 &    2.6 & 17 & 1691.6 &  4.0 $\pm$  1.4 & 4.1 &  8.5 &  2.5 & 3C 390.3 & AGN, Seyfert 1  \\ 
SAXWFC J1845.6+0051.3 & 18 45 34.4 &  00 51 19.8 & 1.4 &    24.0 &    56.2 $\pm$ 2.2 &   28.7 &   32.4 & 11 & 209.4 & 43.3 $\pm$ 14.2 & 42.3 & 64.0 & 13.0 & GS 1843+009 & Be HMXRB, XP, T  \\ 
SAXWFC J1848.2-0225.9 & 18 48 14.4 & -02 25 54.8 & 1.4 &    15.2 &    24.7 $\pm$ 2.4 &   10.3 &   17.7 & 4 & 114.2 & 27.2 $\pm$  5.0 & 1.7 & 36.0 & 25.0 & GS 1843-02 & Be HMXRB, XP, T  \\ 
SAXWFC J1848.3-0310.6 & 18 48 18.0 & -03 10 38.2 & 1.5 &    46.6 &    22.6 $\pm$ 2.4 &    9.3 &    7.4 & 0 & -- & --- & -- & -- & -- & IGR J18483-0311 & unclassified    \\ 
SAXWFC J1853.1-0844.0 & 18 53 03.5 & -08 44 00.6 & 2.4 &     8.2 &    16.1 $\pm$ 2.1 &    7.3 &    1.7 & 1 & -- & --- & -- & -- & -- & 4U 1850-087 & LMXRB, GlC, B  \\ 
SAXWFC J1855.5-0237.1$^{i}$ & 18 55 30.8 & -02 37 08.7 & 2.2 &    43.6 &     7.4 $\pm$ 2.5 &    2.8 &    5.9 & 8 & 335.7 &  6.5 $\pm$  8.6 & 2.7 & 29.0 &  2.3 & XTE J1855-026 & SG HMXRB, XP   \\ 
SAXWFC J1856.8+0518.4 & 18 56 45.0 &  05 18 23.4 & 2.7 &     6.3 &   183.1 $\pm$ 5.8 &   38.9 &    2.3 & 11 & 160.3 & 120.7 $\pm$ 43.8 & 45.7 & 180.0 & 50.0 & XTE J1856+053 & LMXRB, BH, T \\ 
SAXWFC J1858.7+2240.0 & 18 58 42.9 &  22 39 57.2 & 1.3 &    81.0 &  1416.1 $\pm$ 28.9 &  133.7 &   62.0 & 5 & 357.9 & 13.6 $\pm$ 781.7 & 1140.8 & 1500.0 &  8.2 & XTE J1859+226 & LMXRB, BH, T  \\ 
SAXWFC J1908.9+0922.3 & 19 08 51.9 &  09 22 15.9 & 1.3 &     8.4 &    86.4 $\pm$ 3.4 &   27.3 &    0.7 & 3 & 18.5 & 76.2 $\pm$ 10.8 & 7.9 & 86.0 & 66.0 & XTE J1908+094 & LMXRB, BH, T    \\ 
SAXWFC J1909.6+0949.8 & 19 09 36.2 &  09 49 49.4 & 1.4 &    29.1 &    46.1 $\pm$ 2.7 &   16.7 &   10.9 & 19 & 605.7 & 22.4 $\pm$  8.4 & 11.2 & 46.0 & 14.0 & 4U 1907+097 & SG HMXRB, XP   \\ 
SAXWFC J1910.8+0735.6 & 19 10 50.7 &  07 35 33.7 & 1.7 &    29.1 &    15.6 $\pm$ 1.7 &    9.4 &    7.7 & 8 & 319.3 & 16.7 $\pm$  6.1 & 4.6 & 32.0 & 14.0 & 4U 1908+075 & HMXRB, XP   \\ 
SAXWFC J1911.3+0035.5 & 19 11 16.5 &  00 35 27.9 & 1.3 &    48.7 &  1602.9 $\pm$ 31.4 &  177.0 &   69.5 & 10 & 321.3 & 274.7 $\pm$ 497.7 & 662.9 & 1600.0 & 86.0 & Aql X-1 & LMXRB, B, T   \\ 
SAXWFC J1911.8+0459.8 & 19 11 49.5 &  04 59 48.4 & 1.5 &    36.3 &    20.0 $\pm$ 2.1 &    9.1 &    2.8 & 5 & 168.5 & 19.4 $\pm$ 18.5 & 5.7 & 60.0 & 18.0 & SS 433 & SG HMXRB  \\

SAXWFC J1915.2+1057.4 & 19 15 11.0 &  10 57 21.6 & 1.3 &    11.8 &  2565.9 $\pm$ 50.5 &  452.7 &  134.1 & 93 & 2582.0 & 889.1 $\pm$ 957.3 & 603.0 & 4300.0 & 460.0 & GRS 1915+105 & LMXRB, BH, T \\ 
SAXWFC J1918.8-0514.9 & 19 18 45.4 & -05 14 51.3 & 1.4 &    36.3 &    36.8 $\pm$ 1.6 &   25.4 &   10.2 & 21 & 533.3 & 36.7 $\pm$ 17.1 & 33.9 & 77.0 & 16.0 & XB 1916-053 & LMXRB, B, D  \\ 
SAXWFC J1921.1-5841.2 & 19 21 08.9 & -58 41 09.6 & 1.6 &    72.1 &     5.8 $\pm$ 0.7 &    7.9 &    2.7 & 1 & -- & --- & -- & -- & -- & ESO 141-G55 & AGN, Seyfert 1  \\ 
SAXWFC J1921.2+4400.1 & 19 21 14.4 &  44 00 06.8 & 1.6 &    31.6 &    11.5 $\pm$ 1.3 &    8.2 &    0.8 & 9 & 493.0 &  8.9 $\pm$  4.2 & 3.4 & 21.0 &  7.2 & ABELL 2319 & ClG  \\ 

SAXWFC J1924.4+5014.5$^{i}$ & 19 24 26.9 &  50 14 28.6 & 1.8 &    68.7 &     0.0 $\pm$ 0.0 &    0.0 &    0.0 & 0 & -- & --- & -- & -- & -- & CH Cyg & Sy   \\ 
SAXWFC J1928.5-3507.9 & 19 28 30.6 & -35 07 51.2 & 1.5 &    42.3 &    17.6 $\pm$ 1.4 &   12.3 &    5.6 & 1 & -- & --- & -- & -- & -- & HD 182928 & G5V star    \\ 
SAXWFC J1945.7+2721.4 & 19 45 40.0 &  27 21 21.9 & 1.3 &    34.5 &    98.3 $\pm$ 2.4 &   69.1 &   55.4 & 21 & 1161.9 & 35.7 $\pm$ 30.7 & 168.4 & 100.0 & 11.0 & XTE J1946+274 & Be HMXRB, XP, T  \\ 
SAXWFC J1949.6+3011.8 & 19 49 34.2 &  30 11 47.0 & 1.3 &    48.6 &   165.9 $\pm$ 3.8 &   74.3 &   51.8 & 9 & 434.7 & 37.7 $\pm$ 56.6 & 204.1 & 170.0 & 24.0 & KS 1947+300 & Be HMXRB, XP, T    \\ 
SAXWFC J1955.7+3205.3 & 19 55 40.8 &  32 05 16.4 & 1.3 &    83.3 &    64.3 $\pm$ 1.7 &   58.8 &   37.7 & 35 & 1371.5 & 29.1 $\pm$ 21.4 & 80.8 & 110.0 & 12.0 & 4U 1954+319 & HMXRB    \\ 
SAXWFC J1958.4+3513.1 & 19 58 23.9 &  35 13 03.7 & 1.3 &    86.4 &  1162.9 $\pm$ 23.0 &  522.7 &  105.3 & 163 & 5471.8 & 586.3 $\pm$ 434.7 & 252.54 & 2300.0 & 170.0 & Cyg X-1 & SG HMXRB, BH  \\ 
SAXWFC J1959.4+1141.8 & 19 59 22.2 &  11 41 46.3 & 1.3 &    52.5 &    46.5 $\pm$ 1.3 &   48.8 &    3.0 & 70 & 1850.1 & 64.7 $\pm$ 28.1 & 40.3 & 160.0 & 40.0 & 4U 1957+115 & LMXRB, BH  \\ 
SAXWFC J1959.4+4044.0 & 19 59 25.1 &  40 44 01.6 & 1.6 &    57.9 &    10.5 $\pm$ 1.4 &    7.2 &    2.9 & 2 & 109.9 &  9.7 $\pm$  1.0 & 0.5 & 10.0 &  9.0 & Cygnus A & QSO  \\ 
SAXWFC J2000.0+6508.8 & 20 00 01.0 &  65 08 45.2 & 1.4 &    23.5 &    23.6 $\pm$ 0.9 &   28.7 &    5.7 & 42 & 3164.2 &  9.6 $\pm$  4.4 & 18.8 & 25.0 &  4.6 & 1ES 1959+650 & AGN, BL Lac   \\ 
SAXWFC J2008.6-6024.6 & 20 08 35.7 & -60 24 33.8 & 2.1 &    67.6 &    18.0 $\pm$ 2.5 &    6.3 &    0.0 & 1 & -- & --- & -- & -- & -- & 1RXS J200905.6-602537 & WD   \\
SAXWFC J2008.6-6107.4 & 20 08 36.2 & -61 07 23.5 & 2.0 &    91.5 &     4.9 $\pm$ 0.7 &    7.0 &    1.8 & 1 & -- & --- & -- & -- & -- & NGC 6860 & AGN, Seyfert 1  \\ 
SAXWFC J2009.4-4850.0 & 20 09 24.1 & -48 50 00.2 & 1.3 &    70.4 &    32.9 $\pm$ 1.1 &   36.3 &    9.0 & 22 & 1045.1 & 16.2 $\pm$  9.0 & 77.3 & 35.0 &  7.7 & PKS 2005-489 & AGN, BL Lac   \\ 
\hline

\end{tabular}
\end{minipage}
\end{table*}

\newpage
\thispagestyle{empty}
\markright{}
\authorrunning{}
\titlerunning{}

\begin{table*}[tbh]
\setcounter{table}{2}

\scriptsize

\begin{minipage}{250mm}
\caption{ BeppoSAX WFC X-ray Source Catalogue}

\begin{tabular}{rccccrrrrrrlllll}
\hline
\hline
 WFC Name & RA(2000) & DEC(2000) & $^{a}$ERR$_{RAD}$ & EXPO &
   $^{b}$Flux$_{2-10}$ & $^{c}$SNR1 & $^{d}$SNR2 & N$_{det}$ &
   EXPO$_{total}$ & $^{e}<$FLUX$>_{2-10}$ & $\chi^{2}_{flux}$  &
   $^{f}$Flux$_{MAX}$  & $^{f}$Flux$_{MIN }$ & Name & Class   \\
   & hh mm ss & o  '  "   & arcmin & ks~~~ & &  &  &   & ks ~~~~~ &  & &  & &  &
  \\
\hline
SAXWFC J2012.7+3811.0 & 20 12 39.3 &  38 10 57.7 & 1.3 &     3.3 &   289.2 $\pm$ 6.8 &   74.3 &    0.0 & 7 & 200.9 & 41.4 $\pm$ 141.0 & 573.8 & 310.0 & 16.0 & XTE J2012+381 & LMXRB, BH, T \\ 
SAXWFC J2012.7-5650.1 & 20 12 39.9 & -56 50 05.2 & 1.6 &    74.5 &     4.6 $\pm$ 0.6 &    7.3 &    0.0 & 1 & -- & --- & -- & -- & -- & ABELL 3667 & ClG  \\ 
SAXWFC J2032.2+3738.2 & 20 32 13.6 &  37 38 14.6 & 1.4 &    27.9 &    31.8 $\pm$ 1.4 &   25.2 &   15.7 & 6 & 206.7 & 29.1 $\pm$  6.6 & 8.8 & 37.0 & 21.0 & EXO 2030+375 & Be HMXRB, XP, T  \\ 
SAXWFC J2032.5+4057.5 & 20 32 27.0 &  40 57 28.8 & 1.3 &    30.6 &   623.8 $\pm$ 12.4 &  311.2 &   67.3 & 170 & 6062.5 & 269.9 $\pm$ 206.9 & 421.3 & 890.0 & 130.0 & Cyg X-3 & HMXRB  \\ 
SAXWFC J2044.2-1043.6 & 20 44 10.9 & -10 43 35.0 & 1.8 &    85.8 &     6.1 $\pm$ 0.7 &    8.3 &    1.5 & 4 & 151.8 &  7.5 $\pm$  4.9 & 6.2 & 17.0 &  6.1 & MKN 509 & AGN, Seyfert 1  \\ 
SAXWFC J2058.8+4147.6 & 20 58 45.4 &  41 47 36.9 & 1.7 &    12.6 &    22.8 $\pm$ 2.6 &    8.8 &    7.4 & 3 & 59.9 & 18.7 $\pm$  3.9 & 3.4 & 23.0 & 15.0 & GRO J2058+42 & Be HMXRB, XP  \\ 
SAXWFC J2103.6+4545.9 & 21 03 38.1 &  45 45 55.8 & 1.4 &   104.4 &    46.0 $\pm$ 1.9 &   23.3 &    9.4 & 20 & 798.5 & 20.5 $\pm$ 11.6 & 29.5 & 58.0 &  7.2 & SAX J2103.5+4545 & Be HMXRB, XP, T  \\ 
SAXWFC J2130.0+1209.7 & 21 29 59.5 &  12 09 44.6 & 1.3 &    20.1 &    51.9 $\pm$ 1.5 &   45.5 &    4.2 & 30 & 964.7 & 30.7 $\pm$ 11.1 & 39.8 & 54.0 & 17.0 & 4U 2127+119 & LMXRB, B, GlC  \\ 
SAXWFC J2132.0-3342.2 & 21 31 59.9 & -33 42 10.4 & 2.2 &    39.5 &     4.5 $\pm$ 0.6 &    7.4 &    2.3 & 1 & -- & --- & -- & -- & -- & CTS 109 & AGN, Seyfert 1  \\ 
SAXWFC J2142.6+4335.8 & 21 42 37.3 &  43 35 50.2 & 1.7 &    53.6 &    24.6 $\pm$ 1.2 &   22.1 &    7.7 & 19 & 1125.6 & 20.0 $\pm$  5.6 & 8.4 & 31.0 & 12.0 & SS Cyg & CV, DN, U Gem \\ 
SAXWFC J2144.7+3818.8 & 21 44 42.0 &  38 18 46.0 & 1.3 &    22.8 &  1403.9 $\pm$ 28.0 &  576.0 &  106.4 & 145 & 4923.4 & 853.2 $\pm$ 335.9 & 307.9 & 1900.0 & 190.0 & Cyg X-2 & LMXRB, B, D  \\ 
SAXWFC J2146.4-8542.7 & 21 46 21.6 & -85 42 41.4 & 1.5 &    33.4 &    17.1 $\pm$ 1.8 &    9.6 &    2.1 & 0 & -- & --- & -- & -- & -- & 1RXS J214641.1-854303 & unclassified    \\ 
SAXWFC J2148.8+8232.2 & 21 48 47.6 &  82 32 10.6 & 1.5 &   164.1 &     6.6 $\pm$ 0.6 &   11.7 &    6.7 & 0 & -- & --- & -- & -- & -- & SAXWFC J2148.8+8232 & unclassified    \\ 
SAXWFC J2158.8-3013.2 & 21 58 48.0 & -30 13 09.8 & 1.9 &    46.3 &     7.6 $\pm$ 0.8 &    9.4 &    2.7 & 8 & 238.8 &  8.2 $\pm$  6.2 & 14.1 & 24.0 &  5.6 & PKS 2155-304 & AGN, BL Lac   \\ 
SAXWFC J2207.9+5431.9 & 22 07 56.9 &  54 31 52.3 & 1.4 &    82.5 &    26.6 $\pm$ 1.0 &   28.6 &    9.6 & 60 & 2873.9 & 13.9 $\pm$  6.4 & 23.5 & 35.0 &  5.0 & 4U 2206+543 & Be HMXRB, XP  \\ 
SAXWFC J2208.8+4545.4 & 22 08 45.0 &  45 45 26.2 & 2.3 &    68.0 &    29.9 $\pm$ 4.2 &    6.1 &    2.5 & 1 & -- & --- & -- & -- & -- & AR Lac & CV, RSCVn  \\ 
SAXWFC J2209.2-4713.3 & 22 09 14.3 & -47 13 15.9 & 1.7 &    39.5 &     4.5 $\pm$ 0.7 &    6.1 &    0.0 & 1 & -- & --- & -- & -- & -- & NGC 7213 & AGN, Seyfert 1  \\ 
SAXWFC J2214.1+1242.6 & 22 14 06.5 &  12 42 34.5 & 1.8 &    66.6 &     5.4 $\pm$ 0.8 &    6.4 &    2.4 & 1 & -- & --- & -- & -- & -- & RU Peg & CV, DN, U Gem \\ 
SAXWFC J2239.2+6116.4 & 22 39 14.3 &  61 16 26.7 & 1.4 &    13.1 &    33.6 $\pm$ 1.6 &   21.6 &   18.1 & 1 & -- & --- & -- & -- & -- & SAX J2239.3+6116 & Be HMXRB, XP, T \\ 
SAXWFC J2253.1+1649.3 & 22 53 05.4 &  16 49 15.6 & 1.5 &    43.4 &    24.2 $\pm$ 1.7 &   13.3 &    2.5 & 1 & -- & --- & -- & -- & -- & IM Peg & RSCVn  \\ 
SAXWFC J2323.5+5848.1 & 23 23 27.6 &  58 48 03.9 & 1.3 &   138.2 &   109.8 $\pm$ 2.3 &  145.2 &   10.1 & 167 & 7876.7 & 103.4 $\pm$ 13.7 & 21.4 & 150.0 & 59.0 & Cassiopeia A & SNR shell   \\ 
SAXWFC J2331.9+1956.1 & 23 31 53.2 &  19 56 05.6 & 1.9 &    33.7 &    11.0 $\pm$ 1.4 &    7.0 &    0.0 & 1 & -- & --- & -- & -- & -- & EQ Peg & star M3V   \\ 
SAXWFC J2359.2-3038.0 & 23 59 09.5 & -30 37 59.8 & 2.2 &    45.9 &     5.3 $\pm$ 0.6 &    7.9 &    1.2 & 2 & 130.6 &  4.0 $\pm$  1.6 & 7.4 &  5.3 &  3.1 & 1H 2354-315 & AGN, BL Lac   \\ 
\hline  
SAXWFC J0234.3-4347.1$^{k}$ &  02 34 19.2 & -43 47 07.4 & 1.9 & 113.7 & 3.9 $\pm$ 0.5 & 7.8 & 2.3 & 1 & -- & --- & -- & -- & -- &   CC Eri             & RSCVn  \\ 
SAXWFC J0759.8-3845.0$^{k}$ &  07 59 45.0 & -38 44 36.2 & 2.1 & 21.9 & 14.4 $\pm$ 2.1 & 6.0 & 0.0 & 1 & -- & --- & -- & -- & -- &  IGR J07597-3842     &  AGN, Seyfert 1   \\ 
SAXWFC J1510.9+0542.7$^{k}$ &  15 10 57.0 &  05 42 39.2 & 1.8 & 56.0 & 6.7 $\pm$ 0.9 & 7.3 & 0.0 & 2 & 100.8 & 7.1 $\pm$ 0.9 & 0.7 & 7.9 & 6.7 &  ABELL 2029            & ClG   \\ 
SAXWFC J1632.0-4751.5$^{k}$ &  16 32 02.0 & -47 51 29.8 & 1.6 & 9.0 & --- & 0.0 & 0.0 & 1 & -- & --- & -- & -- & -- &  AX J1631.9-4752    &  HMXRB, XP         \\ 
SAXWFC J1709.0-3623.2$^{k}$ &  17 09 00.6 & -36 23 14.2 & 1.8 & 41.8 & --- & 0.0 & 0.0 & 1 & -- & --- & -- & -- & -- &   IGR J17091-3624     & XB  \\ 
SAXWFC J1709.5-2639.2$^{k}$ &  17 09 29.9 & -26 39 11.8 & 2.1 & 93.3 & 52.6 $\pm$ 5.4 & 9.3 & 0.0 & 2 & 102.1 & 63.0 $\pm$ 54 & 27.3 & 129.0 & 52.6 & 1RXS J170930.2-263927   &  LMXRB  \\ 
SAXWFC J1725.1+1153.8$^{k}$ &  17 25 07.2 & 11 53 45.2 & 2.1 & 13.7 & 15.6 $\pm$ 2.2 & 6.4 & 0.0 & 1 & -- & --- & -- & -- & -- & 1H 1720+117   &  AGN, BL Lac  \\ 
SAXWFC J1914.1+0951.7$^{k}$ &  19 14 07.7 &  09 51 42.8 & 2.0 & 6.7 & --- & 0.0 & 0.0 & 1 & -- & --- & -- & -- & -- &   IGR J19140+0951    & HMXRB  \\ 
SAXWFC J2055.4-0021.0$^{k}$ &  20 55 24.6 & -00 20 59.2 & 1.9 & 50.0 & 4.96 $\pm$ 0.73 & 6.5 & 0.0 & 1 & -- & --- & -- & -- & -- & 1RXS J205528.2-002123   & unclassified  \\ 
SAXWFC J2124.7+5058.3$^{k}$ &  21 24 40.8 &  50 58 17.7 & 2.0 & 81.0 & 6.96 $\pm$ 0.83 & 7.8 & 0.0 & 1 & -- & --- & -- & -- & -- & IGR J21247+5058     & AGN, RG  \\ 
SAXWFC J2355.2+2839.4$^{k}$ &  23 55 11.5 &  28 39 23.7 & 1.9 & 34.8 &  --- & 0.0 & 0.0 & 1 & -- & --- & -- & -- & -- &   II Peg           & RSCVn  \\
\hline

\end{tabular}

$^{a}$ the error radius is the mean 99\% X, Y error added in quadrature to a systematic error of 1.3\arcmin         \\
$^{b}$ flux and error in units of 10$^{-11}$\ergcs   \\
$^{c}$ signal-to-noise ratio in the 2-8 keV band     \\
$^{d}$ signal-to-noise ratio in the 8-19 keV band \\
$^{e}$ the flux weighted mean among all source detections and its standard deviation in units of 10$^{-11}$\ergcs \\
$^{f}$ maximum and minimum fluxes in units of 10$^{-11}$\ergcs \\
$^{g}$ Sources classification: AGN, Active galactic  nuclei; AM Her, AM Her type star; AXP, Anomalous X--ray pulsar; Be, B-emission star; B, Burster; BY Dra, BY Dra type star; ClG, Cluster of galaxies; CV, Cataclismic variable; D, dipper; DN, Dwarf nova; E, eclipsing; GlC, in globular cluster; HMXRB, High Mass X--ray Binaries; IP, Intermediate polar; LBV, LBV type star; LMXRB, Low Mass X--ray Binaries; P, Polar; QSO, Quasar; RG, radio galaxy; RP, radio pulsar; RSCVn, RS CVn type star; Seyfert 1 or 2, Seyfert galaxy type 1 or type 2; SNR, Supernova remnant; Sy, Symbiotic star; T, transient source; U Gem, U Gem type star; WD, White dwarf; XB, X--ray binary; XP, X--ray pulsar \\
$^{h}$ detected with the highest snr in a softer band, 2--5.5 or 2--3.2 keV with respect to the catalogue reference 2--8 keV band \\
$^{i}$ detected with the highest snr in a harder band, 8--19, 5.5--11, 6--13 or 11--19 keV with respect to the catalogue reference 2--8 keV band  \\
$^{j}$ detected with the highest snr in a larger, 2--19 or 2--26 keV, band with respect to the catalogue reference 2--8 keV band \\
$^{k}$ source detected in the first processing stage only \\
\end{minipage}
\end{table*}

\end{landscape}


\end{document}